\documentclass[11pt]{emulateapj}
\usepackage{natbib}
\usepackage{graphicx}
\usepackage{amsmath}
\usepackage{amssymb}

\def\Wm2{W/m$^2$}
\def\Wpm2sr{Wm$^{-2}sr^{-1}$}
\def\deg{$^\circ$ }
\def\degx{$^\circ$}

\def\etal{{\it et al.\ }}

\begin{document}

\title{Dynamics of Cloud Features on Uranus\footnotemark[\dag]}
  \author{L.A. Sromovsky\altaffilmark{1} and
    P.\ M. Fry\altaffilmark{1}} \altaffiltext{1}{University of
    Wisconsin - Madison, Madison WI 53706} \altaffiltext{\dag}{Based
    in part on observations with the NASA/ESA Hubble Space Telescope
    obtained at the Space Telescope Science Institute, which is
    operated by the Association of Universities for Research in
    Astronomy, Incorporated under NASA Contract NAS5-26555.}
  
\slugcomment{Journal reference: Icarus 179 (2005) 459-484.}
\begin{abstract}

Near-infrared adaptive optics imaging of Uranus by the Keck 2
telescope during 2003 and 2004 has revealed numerous discrete cloud
features, 70 of which were used to extend the zonal wind profile of
Uranus up to 60\deg N. We confirmed the presence of a north-south
asymmetry in the circulation (Karkoschka, Science 111, 570-572, 1998),
and improved its characterization.  We found no clear indication of
long term change in wind speed between 1986 and 2004, although results
of Hammel et al. (2001, Icarus 153, 229-235) based on 2001 HST and
Keck observations average $\sim$10 m/s less westward than earlier and
later results, and 2003 observations by Hammel et al. (2005, Icarus 175,
534-545) show increased wind speeds near 45\degx N, which we don't
see in our 2003-2004 observations.  We observed a wide range of
lifetimes for discrete cloud features: some features evolve within
$\sim$1 hour, many have persisted at least one month, and one feature
near 34\deg S (termed S34) seems to have persisted for nearly two
decades, a conclusion derived with the help of Voyager 2 and HST
observations.  S34 oscillates in latitude between 32\deg S and
36.5\deg S, with a period of $\sim$1000 days, which may be a result of
a non-barotropic Rossby wave.  It also varied its longitudinal drift
rate between -20\degx/day and -31\degx/day in approximate accord with
the latitudinal gradient in the zonal wind profile, exhibiting
behavior similar to that of the DS2 feature observed on Neptune
(Sromovsky et al., Icarus 105, 110-141, 1993).
S34 also exhibits a superimposed rapid oscillation with an
amplitude of 0.57\deg in latitude and period of 0.7 days, which is
approximately consistent with an inertial oscillation.

\end{abstract}

\maketitle
\shortauthors{Sromovsky and Fry} 
\shorttitle{Dynamics of cloud features on Uranus}

\section{Introduction}

Among the outer planets, Uranus is unique in two key parameters
affecting atmospheric dynamics.  First, it has almost a complete
absence of internal heat flux.  From Voyager IRIS observations Pearl et
al. (1990) estimated the ratio of emission to absorbed sunlight to be
$E$ = 1.06$\pm$0.08, while the corresponding ratio for Neptune is
2.61$\pm$0.28 (Pearl et al. 1991).  This result, which may not be
invariant, does suggest that solar
energy provides the overwhelming drive for atmospheric dynamics on
Uranus.  Second, Uranus' rotational axis is nearly in the plane of its
orbit.  This produces an extreme in the fractional seasonal variation
of local solar flux on Uranus, and thus raises the question of what
seasonal response might be generated.  However, the long radiative
time constant of Uranus, which is longer than its 84-year orbital
period (Flasar et al. 1987), suggests that the response should be
strongly damped. Allison et al. (1991) concluded that at levels below
the visible cloud tops, there would be a thermal phase lag of nearly a
full Uranian season.  Using a radiative-convective model Wallace
(1983) estimated that annual variations would be $\sim$5 K at the
poles and $\sim$0.5 K at the equator. Friedson and Ingersoll (1987)
found that including dynamical heat transports by baroclinic eddies
resulted in peak-to-peak temperature variations that were about half
those estimated by Wallace and in better agreement with IRIS
measurements at the time of the Voyager encounter (Hanel et al. 1986).

The bland appearance of Uranus at the time of the Voyager encounter
seemed consistent with the lack of an internal heat source and the
long radiative time constant.  After months of Voyager 2 imaging, only eight eight
discrete cloud features could be defined well enough to use as tracers
of Uranus' zonal circulation (Smith \etal 1986).  Seven of these were
between 27\degx S and 40\degx S, and one near 70\degx S was visible only
in UV images.  These had relatively low contrast, ranging from from
less than 1\% in violet to about 7\% in the red filter. The features
were also very small, making them difficult to observe in moderate
resolution images.  Subsequent HST imaging at similar wavelengths was
thus understandably not very productive.  The situation improved
significantly in 1997 when the HST NICMOS instrument made possible
near IR imaging of Uranus at high spatial resolution.  Karkoschka
(1998) identified ten new discrete cloud features in the 1997 NICMOS
images and additional features were observed in 1998 images
(Karkoschka and Tomasko 1998). The maximum contrast of northern cloud
features was obtained at 1.87 $\mu$m, reaching 180\% in the raw NICMOS
images.  Near IR imaging at the IRTF in 1998 and 1999 also revealed
discrete cloud features on Uranus (Forsythe et al. 1999; Sromovsky et
al. 2000). These were the first discrete features seen in
groundbased digital images of Uranus, and the 1999 feature was the brightest
ever recorded at 1.7 $\mu$m, comprising $\sim$5\% of the
disk-integrated brightness.    Recent groundbased observations with
the Keck telescope AO system combined exceptional light gathering capabilities
and spatial resolution to reveal a great many more discrete features
(Hammel et al. 2001, de Pater et al. 2002, Hammel et al. 2005),
permitting improved determinations of the Uranus wind profile,
especially at latitudes not previously visible.

Karkoschka (1998) argued that the growing abundance of discrete cloud
features did not represent a significant change in activity of Uranus,
but was due mainly to a change in view angle and improved wavelength
selection.  Although his observations had relatively sparse sampling,
he was able to identify cloud features over long time intervals from
which highly accurate drift rates (or wind speeds) could be defined.
It appeared that cloud features on Uranus had generally very long
lifetimes compared to cloud features on Neptune, few of which can be
found on subsequent rotations (Limaye and Sromovsky 1991). On the
other hand, the Wallace (1983) model predicts seasonal variations in
convective activity, with most of the convection occurring between the
winter solstice and shortly after equinox (2007).  The Friedson and Ingersoll
(1987) model predicts that during the 1994-2007 period, the
temperature contrast between north and south should be near a maximum.
It remains to be understood whether the small temperature variations
these models predict can explain what is turning out to be significant
north-south asymmetries in the dynamics and motions of discrete cloud
features.

New groundbased images we obtained in 2003 and 2004 from the Keck
telescope provide a new abundance of discrete cloud features that we
use to gain a better understanding of Uranus' atmospheric dynamics. In
the following we discuss their morphologies, motions, and evolution,
and include in the analysis results from Voyager and HST observations.
We extend the zonal wind observations up to 60\deg N, and show that
the north-south asymmetry of Uranus' zonal circulation is mainly
confined to mid latitudes. We were able to determine highly accurate
drift rates by tracking features that persisted for at least one
month.  We identify a broad range of evolutionary time scales for
discrete cloud features, and show that one feature probably existed
for two decades or more, and has exhibited regular oscillations in
latitude and longitude, which can plausibly be associated with both
Rossby waves and inertial oscillations.

\section{Observations}

\subsection{Recent Keck Observations}

\begin{figure*}[!htb]\centering
\includegraphics[width=6.4in]{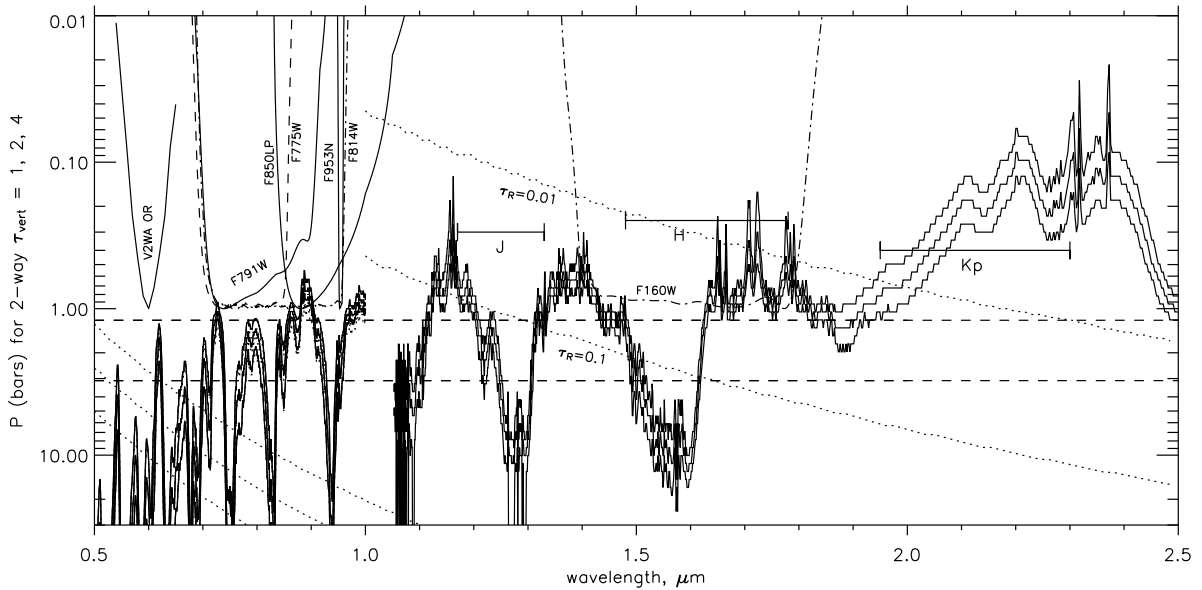}
\caption{Penetration of sunlight into the atmosphere of Uranus vs wavelength. 
Solid curves are shown for two-way vertical optical depths of 1, 2,
and 4 from CH$_4$ and H$_2$ absorption. Dotted curves show the same
optical depths for Rayleigh scattering, except for $\lambda > 1 \mu$m,
where curves are shown for Rayleigh optical depths of 0.1 and 0.01.
Filter passbands for NIRC2 are indicated by horizontal bars. Voyager,
WFPC2, NICMOS, and ACS filters are shown as system throughput curves,
normalized to unity at their peaks. See Table 4 for additional filter
information. Horizontal dashed lines indicate the location of the
putative CH$_4$ cloud at 1.2 bars (Lindal et al. 1987), and the
putative H$_2$S cloud at 3.1 bars (Baines et al. 1995).}
\label{Fig:specpendepth}
\end{figure*}

Using the NIRC2 camera on the Keck 2 telescope we observed Uranus and
Neptune on 15 and 16 July 2003, on 11 and 12 July 2004, and on 11 and
12 August 2004.  The observing geometry for each date is summarized in
Table 1, for 12:00 UT.  Most of our images were made with broadband J,
H, and K' filters using the NIRC2 Narrow Camera. After geometric
correction, the angular scale of this camera is 0.009942$''$/pixel
(NIRC2 General Specifications web page:
http://alamoana.keck.hawaii.edu/inst/nirc2/ genspecs.html). The times,
filters, exposures, and airmass values for these broadband
observations are provided in Tables II and III.  The exposures per
coadd were chosen to avoid saturation of Uranus satellite images so
that they could be used for short-term photometric references.  The
total exposures are less than optimal and represent a compromise
resulting from frequent alternation between Uranus and Neptune imaging
during each observing run.  Images of Hammel et al. (2005) provide
better S/N ratios as a result of longer exposures.  The central
wavelengths and bandwidths of the filters used here are given in Table
4. The pressure levels in Uranus' atmosphere sensed by these filters
is indicated in Fig.\ \ref{Fig:specpendepth}. We also acquired a small
number of narrow-band images in 2004 to provide a cleaner constraint
on vertical cloud structure. These are presented in a separate paper
(Sromovsky and Fry 2005), hereafter referred to as Paper II.

To assess Adaptive Optics (AO) performance, we measured stellar
images, satellite images and epsilon ring cross sections (this ring
is visible in Fig.\ 2). During 15
August 2003 most observations achieved effective seeing of
0.052$''$-0.09$''$ and was mostly in the range 0.061$''$-0.13$''$ on
the following night.  During 11 July 2004 seeing was excellent,
typically 0.05$''$-0.06$''$ with AO turned on.  During 11 August 2004,
raw seeing was about 0.5$''$ in K and effective seeing at 2.15$\mu$m
with AO active reached 0.055$''$.  Natural seeing during 12 August
2004 reached 0.43$''$.  With AO on we measured 0.07$''$-0.08$''$ in H
stellar images (of the relatively faint star FS34). Later measurements
of a brighter star (Elias G 158-27) obtained FWHM values of
0.063$''$-0.070$''$ for H and 0.08$''$-0.090$''$ for J.

\begin{table*}\centering
\caption{Observing geometry for 2003 and 2004 Keck observations.}
\begin{tabular}{|c c c c c c c|}
\hline
         & Range, AU     & Distance, AU& Centric    & Centric   & Phase &\\
     Date (12UT)& (Earth-Uranus)& (Sun-Uranus)& Solar Lat.,\deg & Earth Lat.,\deg& Angle,\deg & MJD\\
\hline
15 AUG 2003 & 19.03 & 20.03 & -16.69 & -16.24 & 0.46 & 52866.5\\
16 AUG 2003 & 19.03 & 20.03 & -16.68 & -16.28 & 0.41 & 52867.5\\
11 JUL 2004 & 19.34 & 20.05 & -13.18 & -11.07 & 2.13 & 53197.5\\
12 JUL 2004 & 19.33 & 20.05 & -13.17 & -11.10 & 2.09 & 53198.5\\
11 AUG 2004 & 19.08 & 20.05 & -12.85 & -12.03 & 0.82 & 53228.5\\
12 AUG 2004 & 19.07 & 20.05 & -12.84 & -12.07 & 0.77 & 53229.5\\
\hline
\end{tabular}
\end{table*}

\begin{table*}\centering
\caption{Uranus observation list for 2003.}
\begin{tabular}{|c c c c| c c c c|}
\hline
Start time& Filter& Itime $\times$coadds& airmass & Start time& Filter& Itime $\times$coadds& airmass\\
\hline
&\multicolumn{2}{c}{15aug03UT}&&&\multicolumn{2}{c}{16aug03UT}&\\
\hline
12:21:05 &         H &  15 $\times$  4 &    1.25 & 10:04:22 &        H &  15 $\times$  4 &    1.21\\
12:22:51 &         H &  15 $\times$  4 &    1.26 & 12:18:23 &        H &  15 $\times$  4 &    1.26\\
12:32:58 &        Kp &  12 $\times$ 10 &    1.28 & 12:22:55 &        J &  15 $\times$  4 &    1.27\\
12:41:25 &         H &  15 $\times$  4 &    1.30 & 12:28:17 &       Kp &  12 $\times$ 20 &    1.28\\
12:48:13 &         H &  15 $\times$  4 &    1.32 & 12:41:51 &        H &  15 $\times$  4 &    1.32\\
12:57:47 &         J &  15 $\times$  4 &    1.36 & 12:46:38 &        J &  15 $\times$  4 &    1.33\\
13:04:22 &         H &  15 $\times$  2 &    1.38 & 12:51:59 &        H &  15 $\times$  4 &    1.35\\
         &           &                 &        & 12:56:16 &         H &  15 $\times$  4 &    1.37\\
         &           &                 &        &13:00:56 &          J &  15 $\times$  4 &    1.38\\
\hline
\end{tabular}
\end{table*}

\begin{table*}\centering
\caption{Uranus observation list for 2004.}
\begin{tabular}{|c c c c| c c c c|}
\hline
Start time& Filter& Itime $\times$coadds& airmass & Start time& Filter& itime $\times$coadds& airmass\\
\hline
&\multicolumn{2}{c}{11jul04UT}& & &\multicolumn{2}{c}{12jul04UT}&  \\
\hline
11:22:15 &         J &  15 $\times$  4 &    1.40 & 10:30:35 &         J &  15 $\times$  4 &    1.69\\
11:30:32 &         H &  15 $\times$  4 &    1.37 & 10:42:40 &         J &  15 $\times$  4 &    1.60\\
11:34:46 &        Kp &  24 $\times$ 10 &    1.35 & 10:49:23 &         H &  15 $\times$  4 &    1.55\\
12:37:57 &         J &  15 $\times$  4 &    1.19 & 11:49:24 &         J &  15 $\times$  4 &    1.29\\
12:42:33 &         H &  15 $\times$  4 &    1.19 & 11:54:22 &        Kp &  24 $\times$ 10 &    1.27\\
13:39:50 &         J &  15 $\times$  4 &    1.15 & 12:05:51 &         H &  15 $\times$  4 &    1.24\\
13:45:22 &         H &  15 $\times$  4 &    1.15 & 13:04:50 &         J &  15 $\times$  4 &    1.16\\
14:13:48 &         H &  15 $\times$  4 &    1.17 & 13:09:47 &        Kp &  24 $\times$ 10 &    1.16\\
14:45:50 &         J &  15 $\times$  4 &    1.21 & 13:36:59 &         H &  15 $\times$  4 &    1.15\\
14:50:35 &         H &  15 $\times$  4 &    1.22 & 14:04:34 &         J &  15 $\times$  4 &    1.16\\
14:54:50 &        Kp &  24 $\times$ 10 &    1.23 & 14:09:18 &         H &  15 $\times$  4 &    1.17\\
15:06:37 &         J &  15 $\times$  4 &    1.25 & 14:38:09 &         J &  15 $\times$  4 &    1.20\\
15:11:09 &         H &  15 $\times$  4 &    1.26 & 14:44:47 &         H &  15 $\times$  4 &    1.21\\
15:15:36 &        Kp &  24 $\times$ 10 &    1.28 & 14:49:05 &        Kp &  24 $\times$ 10 &    1.22\\
15:27:07 &         H &  15 $\times$  4 &    1.31 & 15:00:29 &        Kp &  24 $\times$ 10 &    1.25\\
15:31:27 &         H &  15 $\times$  4 &    1.32 & 15:07:08 &         J &  15 $\times$  4 &    1.26\\
         &           &                 &         & 15:11:42 &         H &  15 $\times$  4 &    1.28\\
         &           &                 &         & 15:16:14 &        Kp &  24 $\times$ 10 &    1.29\\
         &           &                 &         & 15:27:37 &         H &  15 $\times$  4 &    1.33\\
         &           &                 &         & 15:32:23 &         J &  15 $\times$  4 &    1.34\\
\hline
&\multicolumn{2}{c}{11aug04UT} &        &&\multicolumn{2}{c}{12aug04UT} &        \\
\hline
07:11:27 &         J &  15 $\times$  4 &    3.00 & 07:24:55 &         J &  15 $\times$  4 &    2.53\\
07:18:00 &        Kp &  24 $\times$ 10 &    2.80 & 07:29:45 &         H &  15 $\times$  4 &    2.43\\
07:29:45 &         H &  15 $\times$  4 &    2.51 & 07:59:30 &         J &  15 $\times$  4 &    1.96\\
08:08:48 &         J &  15 $\times$  4 &    1.89 & 08:04:32 &        Kp &  24 $\times$  5 &    1.90\\
08:25:04 &         H &  15 $\times$  4 &    1.73 & 08:11:14 &         H &  15 $\times$  4 &    1.83\\
09:30:34 &         J &  15 $\times$  4 &    1.35 & 09:04:37 &         J &  15 $\times$  4 &    1.44\\
09:44:07 &        Kp &  24 $\times$ 10 &    1.30 & 09:09:38 &        Kp &  24 $\times$  5 &    1.42\\
09:53:46 &         H &  15 $\times$  4 &    1.28 & 09:16:53 &         H &  15 $\times$  4 &    1.39\\
10:15:35 &         J &  15 $\times$  4 &    1.23 & 09:46:18 &         J &  15 $\times$  4 &    1.29\\
10:20:30 &         H &  15 $\times$  4 &    1.22 & 09:51:01 &        Kp &  24 $\times$ 10 &    1.27\\
11:32:57 &         J &  15 $\times$  4 &    1.16 & 10:02:39 &         H &  15 $\times$  4 &    1.24\\
11:37:50 &        Kp &  24 $\times$ 10 &    1.16 & 11:11:05 &         J &  15 $\times$  4 &    1.16\\
11:49:33 &         H &  15 $\times$  4 &    1.16 & 11:20:48 &         H &  15 $\times$  4 &    1.16\\
12:45:27 &         J &  15 $\times$  4 &    1.22 & 12:41:43 &         J &  15 $\times$  4 &    1.22\\
12:50:24 &        Kp &  24 $\times$ 10 &    1.23 & 12:46:22 &         H &  15 $\times$  4 &    1.23\\
13:02:11 &         H &  15 $\times$  4 &    1.26 & 13:28:42 &         J &  15 $\times$  4 &    1.36\\
14:02:14 &         J &  15 $\times$  4 &    1.49 & 13:33:25 &        Kp &  24 $\times$ 10 &    1.37\\
14:15:01 &         H &  15 $\times$  4 &    1.57 & 13:45:27 &         H &  15 $\times$  4 &    1.43\\
14:19:37 &        Kp &  24 $\times$  5 &    1.60 & 13:50:42 &         J &  15 $\times$  4 &    1.45\\
         &           &                 &         & 13:55:59 &        Kp &  24 $\times$ 10 &    1.48\\
         &           &                 &         & 14:09:09 &         H &  15 $\times$  4 &    1.55\\
         &           &                 &         & 14:13:51 &         H &  15 $\times$  4 &    1.59\\
\hline
\end{tabular}
\end{table*}

\subsection{Voyager and HST Observations}

The 1986 Voyager 2 and HST images we used for tracking a feature near
34\degx S are listed in Section\ \ref{Sec:char34}. We used
Voyager 2 Wide-Angle orange filter images, with spectral
characteristics given in Table 3 and in Fig.\
\ref{Fig:specpendepth}. The Voyager images were taken at a time when
very little of the northern hemisphere was illuminated.  The HST
WFPC2, NICMOS, and ACS imagery we used for the same purpose are listed
in Section \ref{Sec:char34}. Their filter characteristics
are also given in Table 3 and Fig.\ \ref{Fig:specpendepth}.

\subsection{Image Processing and Navigation}

Our Keck images were dark-subtracted, flat-fielded, geometrically corrected, and
then navigated.  Differences between lamp-on and lamp-off dome image
triplets were median filtered and normalized by the median difference
to define the flat-field correction. 
The significant geometric
distortions in the NIRC2 images were removed using an inversion of the
10-term cubic polynomial given in the NIRC2 pre-ship test report
(Thompson et al. 2001).  
We made use of the SPICELIB toolkit
(Acton 1996) to generate ephemeris information concerning the
orientation of the planet's pole vector, the range to the planet, and
the latitude and longitude of the observer (the point at which a
vector from the planet center to the observer intersects the surface).
We used standard 1-bar radii of $r_{\mathrm{eq}}$ = 25,559 km and
$r_{\mathrm{pol}}$ = 24,973 km and a longitude system based on a
17.24-h period (Siedelmann et al. 2002).  We spot checked image
orientation values in the NIRC2 file header by measuring the positions
of Uranian satellites, and found good agreement.  We determined planet
center coordinates by fitting a projected planet limb to maximum gradient
limb points.
The combined effect of navigation errors and other cloud tracking
errors is approximately 1 narrow camera pixel rms, which is estimated
from the image-to-image scatter found in the plots of discrete cloud
position versus time. In describing positions we use both
planetocentric latitude ($\phi$), which is the angle above the
equatorial plane measured from the planet center, and planetographic
latitude ($\phi_{\mathrm{pg}}$), which is the angle between the local
normal and the equatorial plane. These are related through the
equation $\tan\phi_{\mathrm{pg}}=(r_{\mathrm{eq}}/r_{\mathrm{pol}})^2
\tan\phi$.

HST image processing and navigation proceeded along the lines
described by Sromovsky et al. (2001), except that we used the same
navigation procedure described above.  Voyager 2 Uranus images were
processed using USGS ISIS software (http://isis.astrogeology.usgs.gov)
to correct for vidicon geometric distortions and to remove reseau features.
The images were navigated as described above.

\section{Atmospheric Circulation Results}

Before we present our wind results, it is worth addressing a potential
point of confusion in applying earth-based dynamical arguments to
Uranus.  Defining the rotational pole of Uranus by the right hand rule
places it 98\deg from its orbital pole. Because the tilt puts this
pole south of the invariable plane of the solar system, IAU convention
makes this the south pole of the planet (Seidelmann 2002).  On Uranus,
winds that blow in the same direction as the planet rotates are thus
westward, while on Neptune the prograde winds are eastward. Because
planetographic longitude always increases with time by definition, it
is equal to west longitude on Neptune, but east longitude on Uranus.
To facilitate the application of existing dynamical arguments and
intuition developed from studies of Earth, Jupiter, Saturn, Neptune,
and Mars, it is useful to define dynamically equivalent directions for
Uranus: dynamical north = IAU south and dynamical east = IAU west. We
follow Allison et al. (1991) and Hammel et al. (2001, 2005) in using
prograde (IAU westward or dynamical eastward) winds as positive winds
on Uranus.

\subsection{Cloud-tracked winds from 2003 and 2004}

\begin{figure*}[!hbtp]\centering
\includegraphics[width=6in]{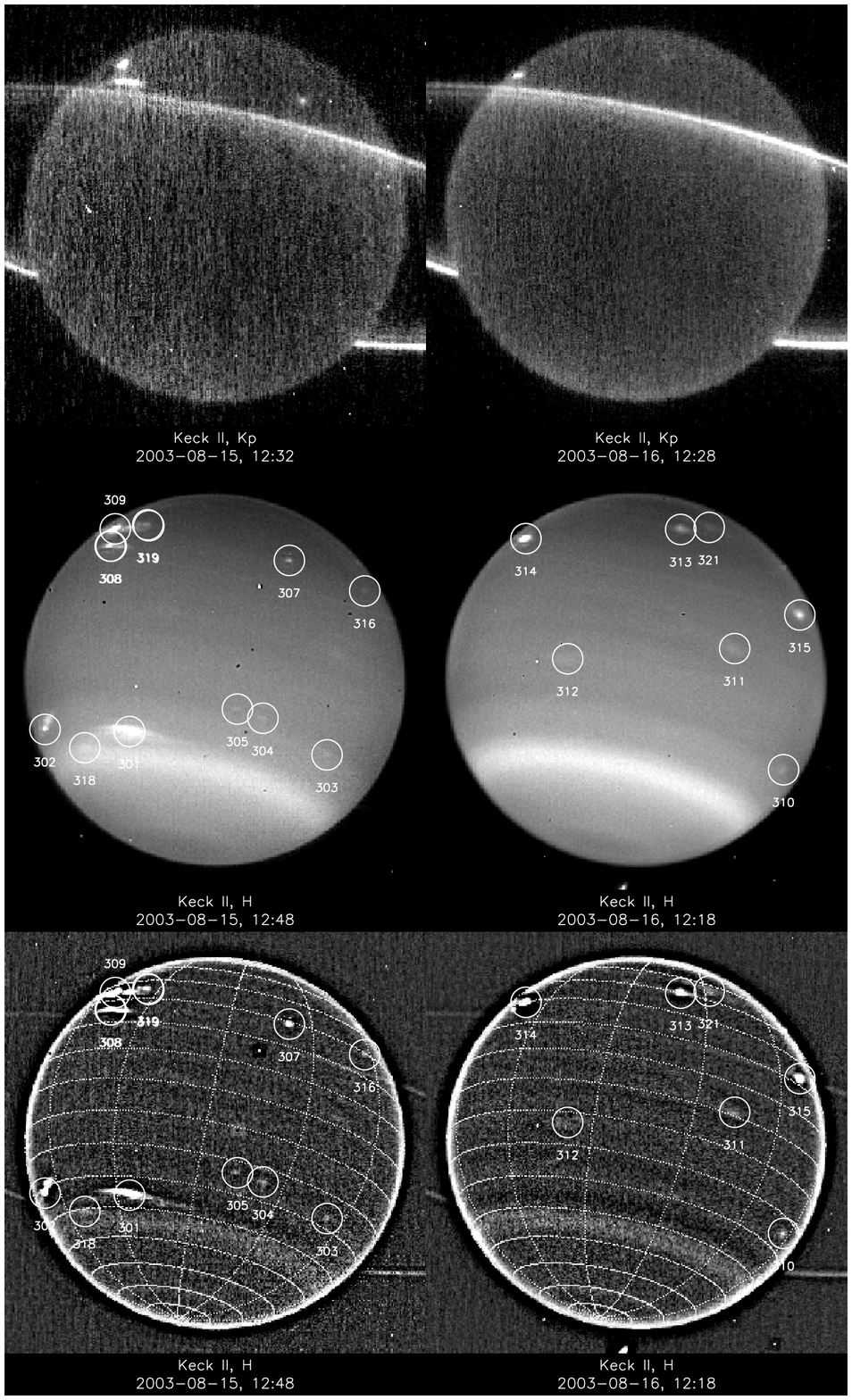}
\caption{Sample images of opposite hemispheres of Uranus in 
August 2003: K$'$ images in upper row, H images in middle row, and
high-pass filtered H images in bottom row, where latitude grid lines
are at 10\deg intervals, and longitude grid lines are at 30\deg
intervals.  Target numbers are used to reference results in Table
5. }
\label{Fig:2003images}
\end{figure*}

\begin{figure*}[!hbtp]\centering
\hspace{0.5in}\includegraphics[width=6in]{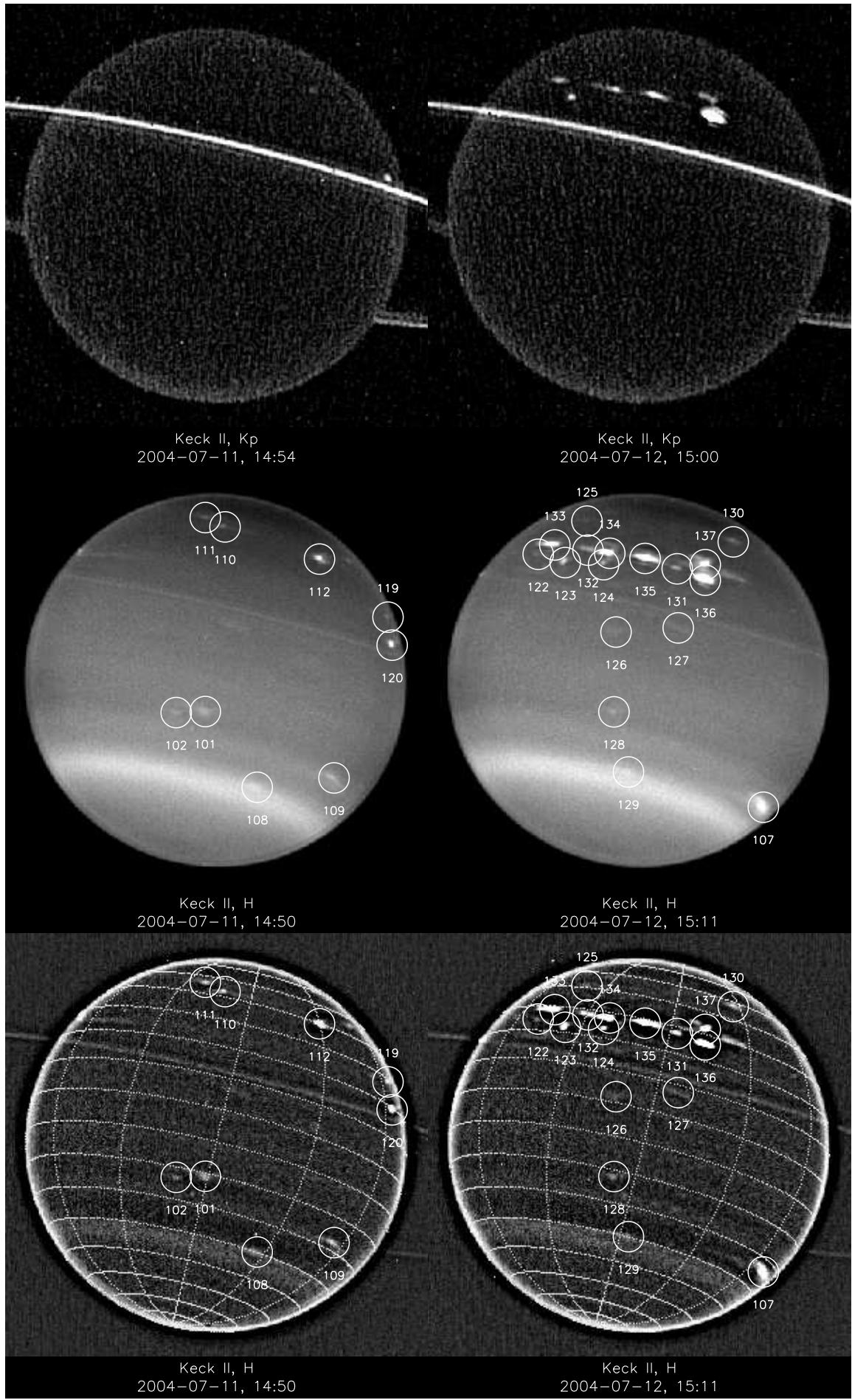}
\caption{Same as Fig.\ \ref{Fig:2003images}, except images are from July 2004
observations.}
\label{Fig:2004aimages}
\end{figure*}

\begin{figure*}[!hbtp]\centering
\hspace{0.5in}\includegraphics[width=6in]{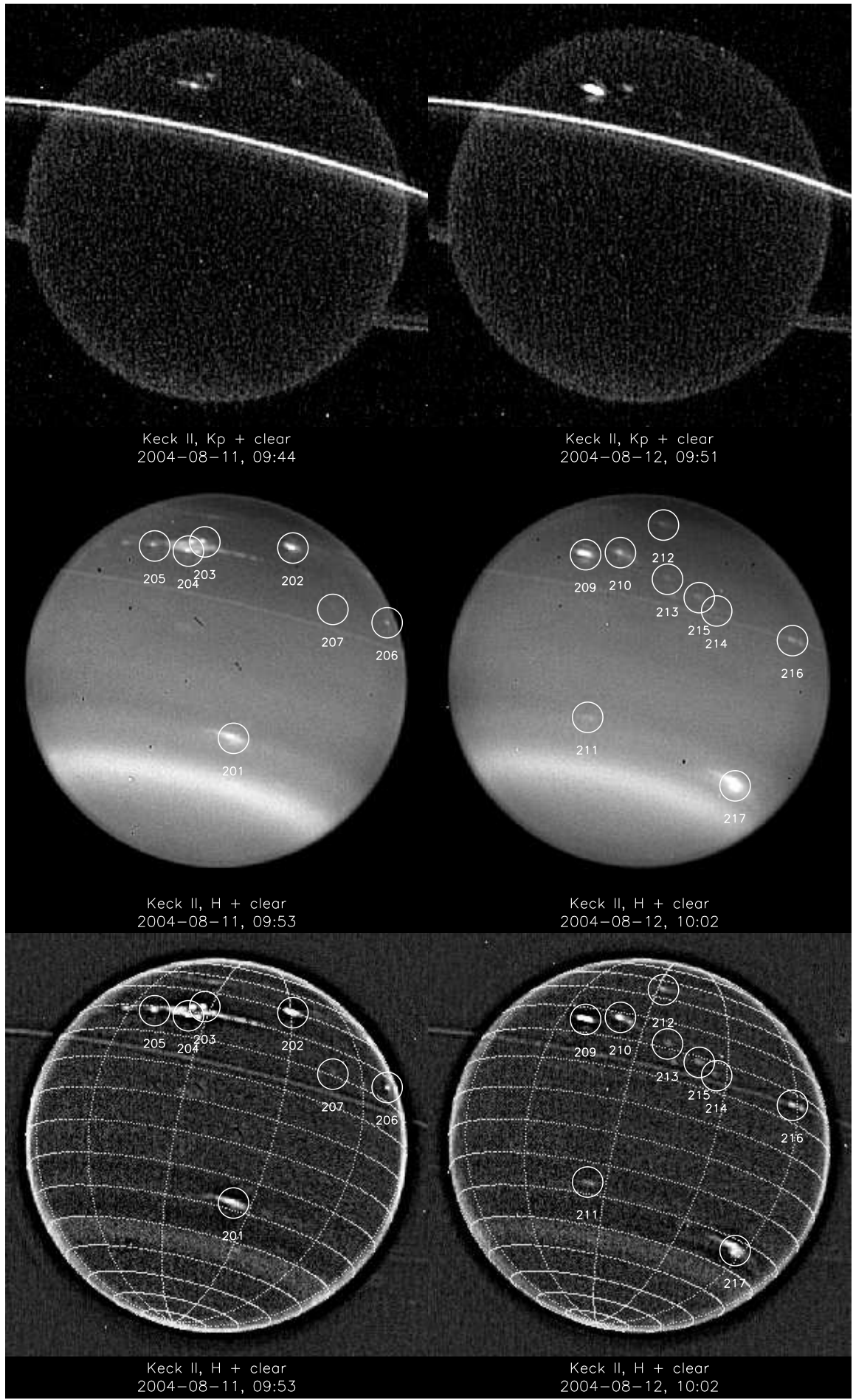}
\caption{Same as Fig.\ \ref{Fig:2003images}, except images are from August 2004
observations.}
\label{Fig:2004bimages}
\end{figure*}

A sampling of the discrete cloud features we used to track
atmospheric motions is provided for each of the three
observing runs in Figs. \ref{Fig:2003images},
\ref{Fig:2004aimages}, and \ref{Fig:2004bimages}.  Images in the top row
of each figure were obtained using the K$'$ filter and the middle row
with the H filter, which provides somewhat better contrast than the J
filter, presumably due to the reduced contribution of Rayleigh
scattering at the longer wavelengths of the H filter (see Fig.\
\ref{Fig:specpendepth}). The bottom row in each figure displays a
high-pass filtered version of the H image to enhance contrast for
subtle cloud features.  All the encircled cloud targets were observed
on multiple images and were used to determine drift rates and wind
speeds. Only the higher altitude cloud features are visible in the
K$'$ image.  In Paper II we show that the highest of these features
reach to $\sim$200 mb.  The northern boundary of the southern bright
band is also a point at which discrete cloud features develop (108 and
129 in Fig.\
\ref{Fig:2004aimages}).  We have also seen discrete clouds at this
boundary in 1998 NICMOS images.  In the equatorial regions, cloud
features are particularly fuzzy and of low contrast (see 306, 312, and
311 in Fig.\ \ref{Fig:2003images}).  Hammel et al. (2004) tracked six
of these features within 2\deg of the equator.  We were able to track
six between 0\degx N and 10\degx N.

Each of our Keck runs provided observations on two successive nights,
between which Uranus undergoes 1.39 rotations.  As a result, there
were only a few cloud targets seen on both nights of each observing
run. These provided the highest wind speed accuracy, of the order of
1.7 to 8 m/s. Most targets were followed during only a single transit
on one night, for which the best accuracy is about 13-15 m/s, for a
target tracked over about 5 hours.  The longest tracking time within a
single transit was 5.5 hours.  These estimated uncertainties in the
zonal winds are of the same order as uncertainties in the meridional winds,
although meridional wind speeds are much smaller.  Because we could
not resolve any systematic meridional motion for features observed on
only one transit, meridional wind speeds are not tabulated.  However, rather
substantial meridional excursions of $\pm$2.1\deg of latitude were
observed for a long-lived feature near 34\degx S, which is discussed
in detail in a Section
\ref{Sec:34move}.  Models imply that this feature regularly reaches peak 
meridional speeds of 28 m/s, but varies so rapidly that
it is difficult to resolve by direct measurement.

\begin{table*}\centering
\caption{Spectral responsivity characteristics for NIRC2, Voyager 2, 
WFPC2, ACS, and NICMOS filter bands we used.}
\begin{tabular}{|c c c c c c|}
\hline
           &         & Central           & Bandpass$^a$, & cut-on$^b$,  & cut-off$^b$, \\
Instrument &  Filter & $\lambda$, $\mu$m & $\mu$m & $\mu$m & $\mu$m\\
\hline
NIRC2$^c$ & J        & 1.248  & 0.163  & 1.166  & 1.330\\ 
NIRC2$^c$ & H        & 1.633  & 0.296  & 1.485  & 1.781\\
NIRC2$^c$ & K$'$       & 2.124  & 0.351  & 1.948  & 2.299\\
Voyager 2$^d$& WA Orange& 0.60 & 0.04 & 0.58 & 0.62\\
  WFPC2$^e$ & 791W  & 0.790 & 0.130  & 0.710       & 0.865     \\
  WFPC2$^e$ & F850LP& 0.909 & 0.106  & 0.85      &  0.956    \\
  WFPC2$^e$ & F953N & 0.954 & 0.005 &      &     \\
      ACS$^f$   & F814W & 0.8333 & 0.251 &0.708 & 0.959 \\
      ACS$^f$   & F775W & 0.7764 & 0.1528 &0.700   &0.853  \\
NICMOS NIC1$^g$ & F160W & 1.603 & 0.399 & 1.40 & 1.80\\
\hline
\end{tabular}\par
\parbox{4.in}{\noindent $^a$full width at half maxima, except for WFPC2 which is effective bandpass.\newline
$^b$half maximum point.\newline
$^c$from www.alamoana.keck/hawaii.edu/inst/nirc2/filters.html.\newline
$^d$from Danielson et al. (1981). \newline
$^e$from WFPC2 Instrument Handbook v 9.0, Oct. 2004.\newline
$^f$from ACS Instrument Handbook v5.0, Oct. 2004.\newline
$^g$from Appendix A, NICMOS Instrument Handbook v 7.0, Oct. 2004, system throughput.}
\end{table*}

\begin{figure}[!hbtp]\centering
\includegraphics[width=3.3in]{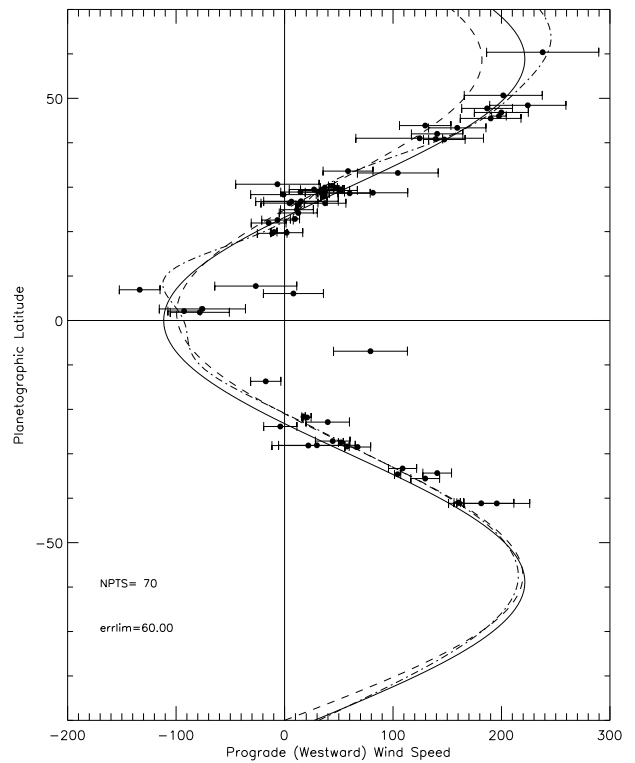}
\caption{Wind observations obtained from Keck imaging in August 2003
  and July and August 2004. The solid line is a symmetric fit to
  Voyager results (Smith et al. 1986) combined with HST results
  (Karkoschka 1998; Hammel et al. 2001). The dashed line is the
  asymmetric fit of Karkoschka (1998). The dot-dash line is our
  asymmetric fit described in the text. Prograde winds are dynamically
  equivalent to eastward winds on Earth.}
\label{Fig:targwinds}
\end{figure}

In Tables 5 and 6, we provide a summary of discrete feature
observations during August 2003 and during July and August 2004.
These are a subset for which wind speed uncertainties are less than 60
m/s.  Most of the errors are less than 40 m/s.  These results are
obtained by fitting individual longitude observations to a linear
drift model, weighting individual observations by the inverse of their
expected variances. The error estimates are computed in the pixel
domain, and taken to be 1 pixel rms. As noted previously, this
estimate incorporates tracking and image navigation errors.  
We made a systematic comparison of expected errors
based on a 1-pixel position uncertainty from all sources with standard
deviations from a linear longitude-vs-time regression line, and found
generally good agreement. The 70 tabulated individual target winds
are plotted in Fig.\
\ref{Fig:targwinds}. The wind speed of 238$\pm$52 m/s at 60\degx N is
for the northernmost and highest speed cloud so far tracked in the
atmosphere of Uranus. However, this target is not visible in the
sample images displayed in Fig.\ \ref{Fig:2004aimages}.  Targets tend
to cluster in narrow latitude bands, leaving gaps near 37\degx N and
especially south of 42\degx S.  The only discrete feature ever tracked
in the latter region was seen only in Voyager 2 UV images (Smith et
al., 1986).  No features were tracked between 60\degx N and the
terminator (73\deg N in 2003 and 77\deg N in 2004), which is at least
partly due to poor viewing geometry.

\begin{table*}\centering
\caption{Individual zonal wind determinations by cloud target (Northern Hemisphere).}
\begin{tabular}{|r r r r l c r c|}
\hline
Lat. ($\phi$) & Lat. ($\phi_{pg}$) & Drift rate (\degx E/h) & 
u(m/s westward) & ID$^a$ & N$_\mathrm{obs}$ & $<\lambda>$(\degx E) & $<$Mod. Julian Day$>$\\
\hline
   59.52&   60.38$\pm$0.54&  -3.82$\pm$0.83& 238$\pm$52&    146&    4&    153.4&   53198.60118\\
   49.70&   50.67$\pm$0.41&  -2.53$\pm$0.45& 202$\pm$36&    148&    3&    337.4&   53197.52743\\
   47.44&   48.42$\pm$0.39&  -2.69$\pm$0.42& 224$\pm$35&    223&    3&     28.9&   53228.48894\\
   46.75&   47.73$\pm$0.27&  -2.21$\pm$0.28& 187$\pm$23&    141&    6&    152.0&   53198.56881\\
   45.84&   46.83$\pm$0.26&  -2.33$\pm$0.29& 200$\pm$25&    111$^1$& 6&     9.7&   53197.59845\\
   45.02&   46.01$\pm$0.34&  -2.27$\pm$0.08& 198$\pm$07&    138&    4&     68.1&   53197.83650\\
   44.49&   45.48$\pm$0.31&  -2.16$\pm$0.32& 190$\pm$28&    212$^1$&  4&  123.4&   53229.40479\\
   42.89&   43.87$\pm$0.23&  -1.44$\pm$0.26& 130$\pm$24&    110&    6&     18.9&   53197.59845\\
   42.37&   43.35$\pm$0.30&  -1.75$\pm$0.29& 159$\pm$26&    208&    4&     43.4&   53228.51515\\
   41.02&   42.00$\pm$0.25&  -1.51$\pm$0.25& 141$\pm$24&    125&    5&    142.3&   53198.58176\\
   40.04&   41.01$\pm$0.22&  -1.32$\pm$0.62& 125$\pm$59&    313&    7&    199.1&   52867.52967\\
   39.86&   40.84$\pm$0.23&  -1.47$\pm$0.07& 139$\pm$06&  112$^2$&  6&     55.2&   53197.75202\\
   39.76&   40.74$\pm$0.22&  -1.55$\pm$0.20& 147$\pm$19&  202$^2$&  7&    360.3&   53228.42366\\
   32.72&   33.63$\pm$0.20&  -0.56$\pm$0.22&  59$\pm$23&    203&    5&    327.7&   53228.38594\\
   32.31&   33.20$\pm$0.20&  -1.00$\pm$0.36& 104$\pm$37&    137&    5&    192.8&   53198.60978\\
   29.81&   30.67$\pm$0.27&   0.06$\pm$0.36&  -7$\pm$38&    314&    4&    137.3&   52867.46734\\
   29.44&   30.29$\pm$0.21&  -0.40$\pm$0.03&  43$\pm$03&    316&    6&    110.7&   52866.97789\\
   29.08&   29.93$\pm$0.21&  -0.34$\pm$0.04&  37$\pm$04&    219&    5&    223.4&   53229.29605\\
   29.02&   29.86$\pm$0.19&  -0.44$\pm$0.02&  47$\pm$02&    224&    6&    255.3&   53229.15336\\
   28.67&   29.51$\pm$0.19&  -0.48$\pm$0.03&  52$\pm$03&    119&    6&     90.4&   53198.06957\\
   28.65&   29.49$\pm$0.20&  -0.25$\pm$0.21&  27$\pm$23&    221&    5&    298.1&   53228.37825\\
   28.59&   29.43$\pm$0.17&  -0.31$\pm$0.18&  34$\pm$19&    205&    6&    310.6&   53228.37367\\
   28.43&   29.27$\pm$0.16&  -0.48$\pm$0.14&  52$\pm$15&    133&    7&    135.9&   53198.57199\\
   28.42&   29.26$\pm$0.17&  -0.34$\pm$0.21&  37$\pm$23&    135&    6&    171.0&   53198.59216\\
   28.08&   28.91$\pm$0.17&  -0.13$\pm$0.17&  15$\pm$18&    204&    6&    323.6&   53228.37367\\
   28.04&   28.87$\pm$0.16&  -0.34$\pm$0.17&  37$\pm$18&    132&    7&    149.9&   53198.58862\\
   27.92&   28.75$\pm$0.23&  -0.75$\pm$0.29&  81$\pm$32&    105&    4&    314.0&   53197.54380\\
   27.87&   28.69$\pm$0.19&  -0.31$\pm$0.03&  34$\pm$03&    210&    5&    112.5&   53229.24259\\
   27.80&   28.62$\pm$0.17&  -0.55$\pm$0.17&  60$\pm$19&    134&    7&    159.2&   53198.57199\\
   27.52&   28.33$\pm$0.19&   0.01$\pm$0.28&  -1$\pm$30&    131&    5&    182.2&   53198.60978\\
   27.07&   27.88$\pm$0.20&  -0.33$\pm$0.02&  36$\pm$03&    118&    5&     79.1&   53198.15980\\
   26.05&   26.84$\pm$0.18&  -0.14$\pm$0.31&  15$\pm$34&  136$^3$&  5&    192.4&   53198.60978\\
   25.97&   26.76$\pm$0.24&  -0.06$\pm$0.30&   6$\pm$33&    106&    3&    318.8&   53197.52743\\
   25.64&   26.42$\pm$0.22&  -0.04$\pm$0.24&   5$\pm$26&    206&    4&     47.1&   53228.46979\\
   25.61&   26.39$\pm$0.21&  -0.34$\pm$0.17&  37$\pm$19&    145&    4&    118.3&   53198.53421\\
   25.25&   26.02$\pm$0.16&  -0.11$\pm$0.02&  13$\pm$02&  209$^3$&  7&    100.4&   53229.15264\\
   24.19&   24.94$\pm$0.16&  -0.10$\pm$0.13&  11$\pm$15&    122&    6&    131.0&   53198.55996\\
   23.47&   24.20$\pm$0.16&  -0.11$\pm$0.15&  13$\pm$18&    124&    7&    158.2&   53198.57199\\
   22.16&   22.85$\pm$0.16&  -0.08$\pm$0.04&  10$\pm$04&    213&    6&    129.9&   53229.29086\\
   21.91&   22.60$\pm$0.15&   0.06$\pm$0.13&  -6$\pm$15&    123&    7&    143.9&   53198.57199\\
   21.27&   21.94$\pm$0.17&   0.13$\pm$0.14& -15$\pm$16&    207&    5&     16.8&   53228.49459\\
   19.29&   19.91$\pm$0.15&   0.08$\pm$0.02& -10$\pm$03&    120&    8&     95.7&   53198.07502\\
   19.14&   19.76$\pm$0.15&  -0.02$\pm$0.13&   2$\pm$15&    149&    6&    137.8&   53198.55996\\
   19.00&   19.61$\pm$0.14&   0.10$\pm$0.12& -11$\pm$14&    144&    7&    138.3&   53198.57199\\
    7.49&    7.75$\pm$0.17&   0.22$\pm$0.31& -27$\pm$38&    127&    4&    185.6&   53198.60118\\
    6.68&    6.92$\pm$0.13&   1.09$\pm$0.15&-134$\pm$19&    139&    7&     31.7&   53197.60494\\
    5.86&    6.07$\pm$0.19&  -0.07$\pm$0.22&   8$\pm$28&    104&    3&    330.1&   53197.52743\\
    2.51&    2.60$\pm$0.18&   0.61$\pm$0.32& -76$\pm$40&    218&    3&    205.7&   53229.56505\\
    2.04&    2.11$\pm$0.14&   0.75$\pm$0.12& -93$\pm$15&    117&    5&    129.5&   53198.54533\\
    1.77&    1.84$\pm$0.14&   0.63$\pm$0.22& -78$\pm$27&    126&    5&    168.0&   53198.60978\\
\hline
\end{tabular}\newline
\parbox{6in}{$^a$ Superscripts identify corresponding features in July and August images.}
\end{table*}

\begin{table*}\centering
\caption{Individual zonal wind determinations by cloud target (Southern Hemisphere).}
\begin{tabular}{|c c c c l c r c|}
\hline
Lat. ($\phi$) & Lat. ($\phi_{pg}$) & Drift rate (\degx E/h) & 
u(m/s westward) & ID$^a$ & N$_\mathrm{obs}$ & $<\lambda>$(\degx E) & $<$Mod. Julian Day$>$\\
\hline
   -6.69&   -6.92$\pm$0.18&  -0.64$\pm$0.28&  79$\pm$34&    103&    3&    309.1&   53197.52743\\
  -13.25&  -13.69$\pm$0.13&   0.14$\pm$0.12& -17$\pm$14&    312&    6&    171.3&   52867.49295\\
  -21.02&  -21.69$\pm$0.13&  -0.15$\pm$0.01&  17$\pm$02&    222&    6&    256.9&   53228.95719\\
  -21.15&  -21.82$\pm$0.11&  -0.18$\pm$0.03&  21$\pm$04&    101&    8&     29.1&   53197.69015\\
  -22.17&  -22.87$\pm$0.13&  -0.35$\pm$0.17&  40$\pm$20&    128&    6&    173.3&   53198.59216\\
  -23.14&  -23.86$\pm$0.12&   0.03$\pm$0.13&  -4$\pm$15&    102&    7&     20.1&   53197.58146\\
  -26.35&  -27.14$\pm$0.14&  -0.40$\pm$0.14&  44$\pm$16&    211&    5&    114.2&   53229.38630\\
  -26.64&  -27.44$\pm$0.13&  -0.47$\pm$0.02&  52$\pm$02&    303&    6&    108.7&   52866.82704\\
  -27.28&  -28.09$\pm$0.17&  -0.27$\pm$0.32&  30$\pm$35&    143&    4&    178.2&   53198.60118\\
  -27.33&  -28.15$\pm$0.16&  -0.20$\pm$0.31&  22$\pm$34&    142&    4&    197.2&   53198.60862\\
  -27.54&  -28.35$\pm$0.11&  -0.53$\pm$0.02&  58$\pm$02&  109$^4$&  8&     78.8&   53198.05767\\
  -27.66&  -28.48$\pm$0.11&  -0.62$\pm$0.11&  67$\pm$12&  201$^4$&  8&    353.9&   53228.40974\\
  -32.40&  -33.30$\pm$0.13&  -1.04$\pm$0.12& 109$\pm$13&    226&    6&    173.2&   53229.49553\\
  -33.43&  -34.35$\pm$0.13&  -1.37$\pm$0.13& 141$\pm$13&  217$^5$&  6&    173.7&   53229.49553\\
  -33.70&  -34.62$\pm$0.13&  -1.01$\pm$0.03& 104$\pm$03&  107$^5$&  6&    270.8&   53197.88258\\
  -34.64&  -35.57$\pm$0.13&  -1.28$\pm$0.13& 130$\pm$13&    225&    6&    174.0&   53229.49553\\
  -40.03&  -41.01$\pm$0.14&  -1.70$\pm$0.05& 161$\pm$04&    108&    6&     51.1&   53197.75202\\
  -40.05&  -41.03$\pm$0.12&  -1.70$\pm$0.02& 160$\pm$02&    220&    8&    244.1&   53228.96836\\
  -40.17&  -41.15$\pm$0.16&  -1.92$\pm$0.32& 181$\pm$30&    129&    5&    184.0&   53198.60978\\
  -40.20&  -41.17$\pm$0.16&  -2.08$\pm$0.32& 196$\pm$30&    147&    5&    184.5&   53198.60978\\
\hline
\end{tabular}\newline
\parbox{6in}{$^a$ Superscripts identify corresponding features in July and August images.}
\end{table*}

\begin{figure*}[!hbtp]\centering
\includegraphics[width=5in]{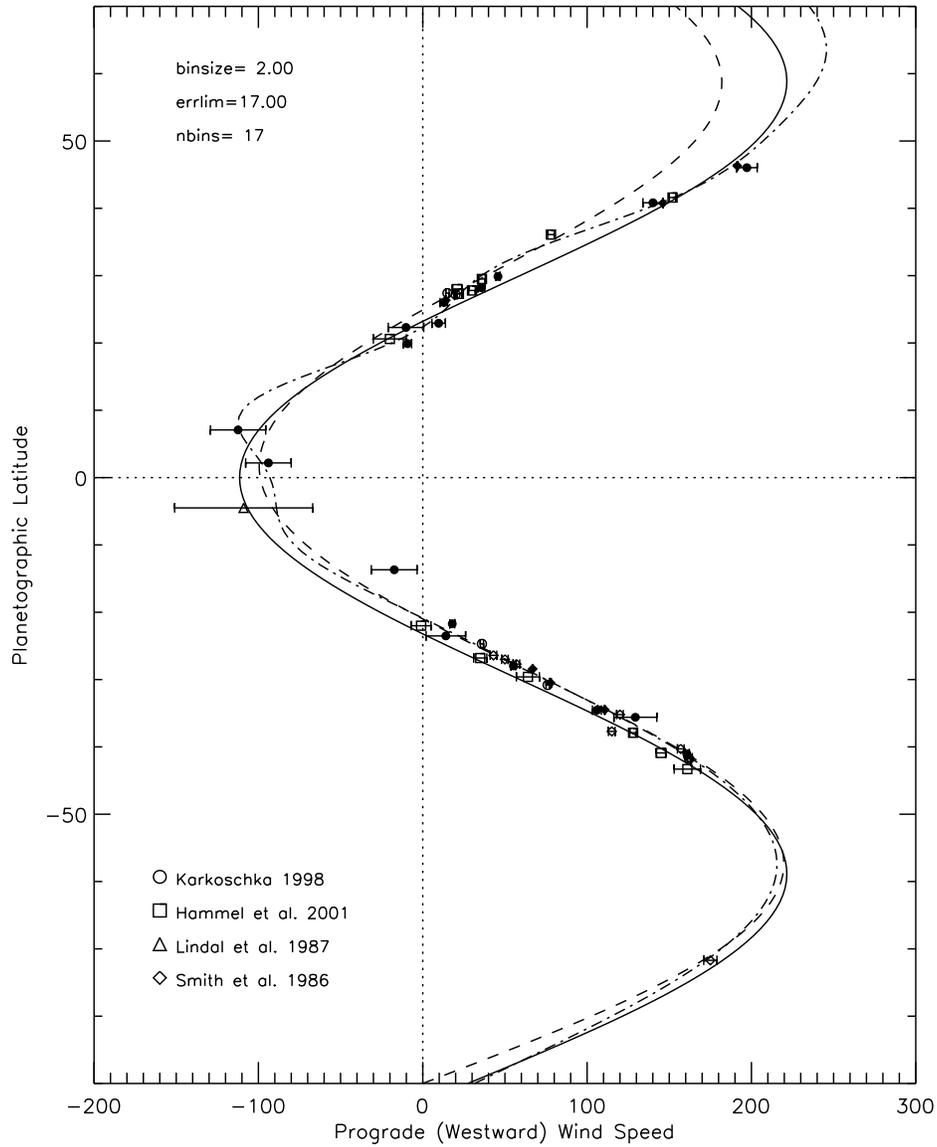}
\caption{Uranus wind observations: binned values with estimated error no greater
than 20 m/s (filled circles). The solid curve is the symmetric fit
given in Fig.\ \ref{Fig:targwinds}. The dashed curve is the asymmetric
fit of Karkoschka (1998), and the dot-dash curve is an asymmetric fit
described int the text. Filled diamonds show our high-accuracy drift
rate results from a month-long time base.}
\label{Fig:windbin}
\end{figure*}

To improve wind speed accuracy in latitude bands with a high density
of cloud targets, we binned results into 2\deg latitude bins and
computed weighted averages.  The binned results are listed in Table
7 and plotted in Fig.\ \ref{Fig:windbin}, where we also show the
prior wind observations of Smith et al. (1986) using Voyager images,
radio occultation results of Lindal et al. (1987), 1997 HST results by
Karkoschka (1998), and results by Hammel et al. (2001) that were
derived from a combination of 1998 HST NICMOS images and 2000 HST
WFPC2 and Keck images. We also include high-accuracy results (shown as
filled diamonds) that we obtained by tracking a select group of clouds
over longer intervals (discussed in Section\ \ref{Sec:lifetime}).  The
solid curve shown in this figure is a symmetric fit to Voyager and HST
observations prior to 2000, given by $u$(m/s) = $27.46 + 36.568
\cos(\phi_{pg}) -175.486 \cos(3\phi_{pg})$.  The dashed curve is a
conversion to wind speed of the asymmetric fit given by Karkoschka
(1998), which expresses the absolute rotation rate in \degx/day as
$482-8\sin\phi_{pg}+127\sin^2\phi_{pg}$ (the planetary rotation of
501.16\deg per day must be subtracted to get relative motions).  The
dot-dash curve is an asymmetric fit to a combination of our Keck
observations and Voyager and Karkoschka HST results, given by $u$(m/s)
$= 62 - 166.5\cos(0.052(\phi_{pg}-2.9)) -20\sin(0.27(\phi_{pg}-27.5))
\times \exp(-(|\phi_{pg}-17.5|/26)^2)+15\cos(0.026(\phi_{pg}-60))$,
for $\phi_{pg}$ in degrees to obtain sine and cosine arguments in
radians.  These data display a clear asymmetry between hemispheres,
which is most noticeable at mid latitudes.  The Hammel et al. (2001)
observations are $\sim$10 m/s less westward than the trend followed by
the other observations.  This is most obvious in the southern
hemisphere.  New observations obtained in 2000 and 2004 at high
northern latitudes are close to the symmetric fit and exceed
Karkoschka's asymmetric fit by $\sim$40 m/s.

\begin{figure}[!hbtp]\centering
\includegraphics[width=3.2in]{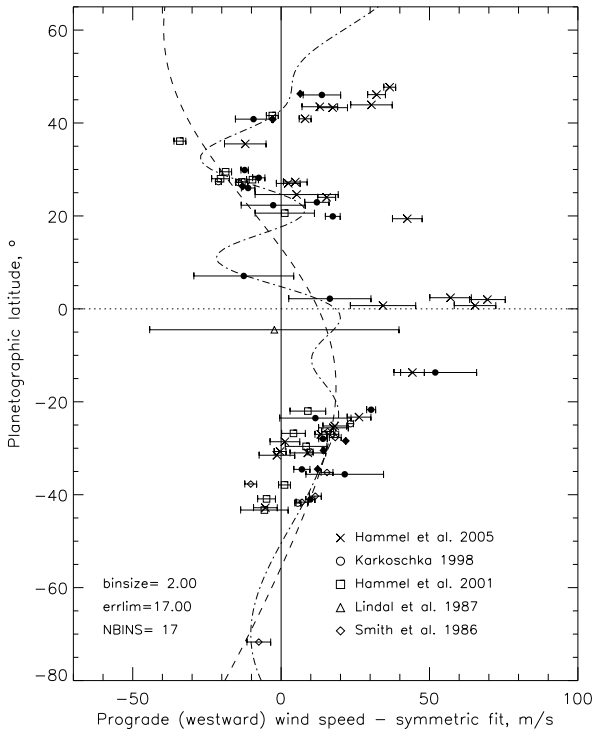}
\caption{Uranus wind observations relative to the symmetric fit given in Fig.\ \ref{Fig:windbin}.
The dashed curve is the difference between the asymmetric
Karkoschka (1988) fit and the symmetric fit. The dot-dash curve is our new
asymmetric fit (see text) minus the symmetric fit. Binned
observations (filled circles) were filtered to include only those with
estimated errors less than 17 m/s. Our high-accuracy winds based on
longer-term tracking are plotted as filled diamonds. Prior measurements are plotted with
symbols defined within the figure. }
\label{Fig:winddiff}
\end{figure}

The comparisons of different wind results is more complete and are more
easily discerned in the difference plot of Fig.\ \ref{Fig:winddiff},
where wind speeds are plotted relative to the symmetric fit shown in
Fig.\ \ref{Fig:windbin}.  Here we also include the more recent observations
of Hammel et al. (2005), which are derived from October 2003 Keck observations.

\begin{table*}\centering
\caption{Binned zonal wind results.}
\begin{tabular}{|r r c c r c r|}
\hline
Lat. ($\phi$) & Lat. ($\phi_{pg}$) & WtdLatDev(\degx) & drift rate (\degx E/h) & u(m/s westward) &  Period (h) & NBIN\\
\hline
   59.52&   60.38$\pm$0.54&  0.00& -4.11$\pm$0.83&  238$\pm$52&     14.40$\pm$0.48&    1\\
   49.70&   50.67$\pm$0.41&  0.00& -2.48$\pm$0.45&  202$\pm$36&     15.41$\pm$0.30&    1\\
   46.97&   47.96$\pm$0.31&  0.32& -2.43$\pm$0.23&  198$\pm$19&     15.45$\pm$0.15&    3\\
   45.05&   46.04$\pm$0.33&  0.24& -2.26$\pm$0.07&  197$\pm$06&     15.55$\pm$0.05&    5\\
   42.66&   43.64$\pm$0.26&  0.26& -1.61$\pm$0.19&  143$\pm$18&     16.01$\pm$0.14&    2\\
   40.89&   41.86$\pm$0.24&  0.34& -1.50$\pm$0.24&  138$\pm$22&     16.09$\pm$0.17&    2\\
   39.85&   40.83$\pm$0.23&  0.03& -1.47$\pm$0.06&  140$\pm$06&     16.11$\pm$0.05&    2\\
   32.61&   33.51$\pm$0.20&  0.19& -0.71$\pm$0.19&   71$\pm$20&     16.67$\pm$0.15&    2\\
   29.03&   29.88$\pm$0.20&  0.26& -0.43$\pm$0.01&   46$\pm$01&     16.89$\pm$0.01&   12\\
   27.39&   28.20$\pm$0.20&  0.40& -0.35$\pm$0.02&   36$\pm$02&     16.96$\pm$0.02&    6\\
   25.24&   26.01$\pm$0.16&  0.18& -0.11$\pm$0.02&   13$\pm$02&     17.15$\pm$0.02&    6\\
   22.23&   22.93$\pm$0.16&  0.30& -0.07$\pm$0.04&   10$\pm$04&     17.18$\pm$0.03&    3\\
   21.62&   22.31$\pm$0.16&  0.32&  0.11$\pm$0.09&  -10$\pm$11&     17.33$\pm$0.08&    2\\
   19.28&   19.90$\pm$0.15&  0.06&  0.09$\pm$0.02&   -9$\pm$03&     17.31$\pm$0.02&    4\\
    6.84&    7.08$\pm$0.13&  0.32&  0.86$\pm$0.14& -112$\pm$17&     17.98$\pm$0.12&    2\\
    5.86&    6.07$\pm$0.19&  0.00&  0.11$\pm$0.22&    8$\pm$28&     17.33$\pm$0.19&    1\\
    2.11&    2.18$\pm$0.15&  0.17&  0.72$\pm$0.11&  -94$\pm$14&     17.86$\pm$0.10&    3\\
    1.70&    1.76$\pm$0.14&  0.21&  0.71$\pm$0.21&  -87$\pm$25&     17.85$\pm$0.18&    2\\
   -6.69&   -6.92$\pm$0.18&  0.00& -0.70$\pm$0.28&   79$\pm$34&     16.68$\pm$0.21&    1\\
  -13.25&  -13.69$\pm$0.13&  0.00&  0.14$\pm$0.12&  -17$\pm$14&     17.35$\pm$0.10&    1\\
  -21.04&  -21.71$\pm$0.13&  0.05& -0.15$\pm$0.01&   18$\pm$02&     17.12$\pm$0.01&    2\\
  -22.80&  -23.51$\pm$0.12&  0.47& -0.09$\pm$0.11&   14$\pm$12&     17.17$\pm$0.09&    3\\
  -24.11&  -24.85$\pm$0.13&  0.01& -0.04$\pm$0.52&    6$\pm$59&     17.20$\pm$0.43&    2\\
  -27.15&  -27.96$\pm$0.12&  0.45& -0.49$\pm$0.01&   55$\pm$01&     16.84$\pm$0.01&    6\\
  -33.63&  -34.55$\pm$0.13&  0.27& -1.04$\pm$0.03&  106$\pm$03&     16.42$\pm$0.02&    3\\
  -34.67&  -35.60$\pm$0.14&  0.19& -1.33$\pm$0.13&  129$\pm$13&     16.21$\pm$0.09&    2\\
  -40.05&  -41.02$\pm$0.13&  0.02& -1.69$\pm$0.02&  161$\pm$02&     15.95$\pm$0.01&    5\\
\hline
\end{tabular}
\end{table*}

There is a considerable dispersion in the measured values for near
equatorial wind speeds.  This may mean that there are wave phenomena
that do not move with the atmospheric mass flow.  Hammel et al. (2004)
suggest a Kelvin wave as a possibility based on the direction of
propagation of equatorial features relative to the zonal flow given by
the Lindal et al. (1987) radio occultation result.  While it is
correct to conclude that the Kelvin wave is more consistent, the
reasoning needs clarification because of the planet's retrograde
rotation. The atmosphere near the equator is actually moving eastward
(dynamically westward) relative to the interior (the period at the
equator is longer than 17.24 hours).  The equatorial clouds that are
suspected of production by wave effects are moving westward
(dynamically eastward) relative to that flow. While Kelvin waves on
earth would move eastward relative to the zonal wind, on Uranus the
direction is reversed and thus consistent with the Hammel et
al. conclusion.  (The directions described in meteorology texts
discussing these waves can be used if one treats Uranus' south pole as
the dynamical north pole of a planet with prograde rotation, and then
treats IAU west coordinates as dynamical east coordinates.)

\subsection{Tracking of long-lived cloud features}\label{Sec:lifetime}

Karkoschka (1998) measured rotation periods for seven cloud features
observed in near-IR HST NICMOS images captured during July and October
1997.  He noted that all seven ``were observed whenever they were on
the illuminated side of the disk'', with ``no evidence for appearance
or disappearance of a cloud during the entire 100-day observation
interval.''  While our recent Keck observations
show that not all Uranus cloud features share this persistence,
we did that find several features live long enough to be seen not only after a
rotation, but also after a month, and even after a year; and one
feature is apparently persistent for decades.

Several features can be reliably identified in both July and August
2004 images based on latitude, morphology, vertical structure, and
consistency of longitudinal drift.  The vertical structure information
is mainly relative brightness in the K$'$ images.  The longitudinal
drift consistency requires observations of a feature on successive
nights during either the July or August observing runs.  From the
drift established in one month, we can project a position in the other
month and compare that with the candidate clouds that are observed
then.

Targets 107, 217, and 301 are easily identified as a single feature
because of similar prominence, morphology, and the lack of competitive
features in the 30\degx S - 40\degx S latitude range.  The
identification is further strengthened by tracking the longitudinal
position.  Because it was seen (as 107) on two successive days in
July, we were able to project a position in August within
$\sim$20\degx. Combining July and August 2004 observations to further
refine the drift rate, we obtain a drift rate with an uncertainty of
only 0.011\degx /day (Table 8), which allows us to predict where
301 should have been in August 2003 with an accuracy of $\sim$3\degx ,
assuming that the drift rate was uniform for the entire year.  The
fact that our projection is off by 100\deg of longitude is evidence
that a uniform drift rate was not present. A detailed discussion of
this feature's dynamics is presented in Section \ref{Sec:34move}.

\begin{table*}\centering
\caption{Medium range tracking summary.}
\begin{tabular}{|c c r r l|}
\hline
            &          &  long. drift rate & Possible ID & Estimated \\
Feature ID  & $<\phi>^a$ &  $d\lambda/dt$,\degx E/\ day & at other times&lifetime\\
\hline
 111 + 212  &  45.31$\pm$0.25 & -53.019$\pm$0.011 & 321? & $\geq 1$ month\\
 112 + 202  &  39.79$\pm$0.09 & -36.999$\pm$0.011 & 350? & $\geq 1$ month\\
 136 + 209  &  25.58$\pm$0.14 & -3.011$\pm$0.008, &308, 309?, 314? & $\geq 1$ month\\
 109 + 201  & -27.60$\pm$0.07 & -14.654$\pm$0.008 & 303? & $\geq 1$ month\\
 302 + ACS  & -31.81$\pm$0.20$^b$&  -17.370$\pm$0.100 &     & $\geq 10$ days \\
 107 + 217  & -33.57$\pm$0.18 &  -25.839$\pm$0.011 & 301 & $\geq 1$ year$^c$\\
\hline
\end{tabular}
\parbox{4.in}{$^a$planetocentric latitude.\newline
$^b$This is the latitude of the brightest element of a 2-component
feature; the mean latitude is -34.21$\pm$0.26\degx.\newline
$^c$This is from Keck only; $\geq 18$ years is estimated from all observations.}
\end{table*}

Targets 209 and 136 are both prominent in K$'$ images, within 0.5\deg
of latitude, and the projected longitude from the August observations
to the July observations is within 8\deg of the observed position,
well within the nominal 15\deg projection error.  Using the July
coordinates to further refine the fit, we obtain a drift rate of
-3.01$\pm$0.01\degx/h at an average latitude of 25.58$\pm$0.14\deg
planetocentric (26.63\deg graphic). The drift rate accuracy is so high
that a position projection back to August 2003 can be made with a
nominal error of $\sim$3\degx.  However, no feature appears very close
to the projected position.  Target 314 is within 30\deg of longitude,
but is at a considerably different latitude ($\sim$30\deg instead of
26\degx).  Feature 308, which is at the expected latitude, is off by
100\deg in longitude.  Feature 307, which is within 30\deg of
longitude, is near 29\deg latitude (3\degx N of the expected position)
and not particularly prominent.  Feature 309, which is the most
prominent, is perhaps the most appealing match from a vertical
structure and morphological point of view, but its position deviates
in both latitude (by $\sim$9\degx) and longitude (by $\sim$100\degx).
Either the feature developed after 2003, or it moved in latitude, in
which case it would likely also have varied its drift rate, and thus is not a clear
match to any 2003 feature. 

Targets 112 and 202 at 39.8\degx N can be convincingly identified as the
same feature.   In this case the projected position in the 2003 August 
images roughly matches a feature seen only near the limb.  But it is
off the projected position by $\sim$50\deg of longitude and $\sim$2\deg
of latitude.  Thus it is clear that the feature cannot have remained
at a fixed latitude and drift rate for the entire year.

HST observations of Uranus with the Advanced Camera for Surveys (ACS) were
made during July and August 2003 (E. Karkoschka, PI).  Images taken on
12 July and 30 August contain the 34\degx S feature, positions for which
are included in Table 9, and discussed in Section \ref{Sec:char34}.  The 30 August
ACS images also contain three prominent northern bright cloud features, two of
which are located at 30\degx N. One of these may be feature 314.  We don't have
sufficient projection accuracy from other observations to determine which
of these, separated by only 12\deg of longitude, is the best candidate to have
been feature 314 on 16 August 2003. Recent HST snapshot observations (K. Rages, PI)
also show features at 30\degx N, one on 29 August 2003 using the F850LP filter,
and another on 5 November 2004 using the F791W filter. 

The ACS images do provide new evidence for the existence of feature
302 in an image taken on 25 August 2003, 10 days after its appearance
in 15 August 2003 images.  The identification is based on unique
2-component morphology in which the two features straddle the 30\degx
S latitude line, with a separation of 4.4\deg in latitude.  The drift
rate computed for this complex is -17.37$\pm$0.10\degx/day, which roughly
matches the Karkoschka (1998) drift rate fit at the mean latitude of the
complex. 

In summary, we were able to track 4 features over a 1-month time interval
during 2004, 1 feature over a 10-day period in 2003, using both ACS and Keck
imagery, and 1 feature over a 1-year period in  Keck imagery, although
the latter feature did not display a uniform drift rate.

\subsection{Evidence for temporal changes in zonal circulation}

Figure\ \ref{Fig:winddiff} displays an apparent time variation in wind
speed.  Southern hemisphere winds in 1986, 1997, and 2004 are
reasonably consistent with each other, but in 2000 and during 2003
winds are less westward (less prograde) by $\sim$10 m/s, especially between 30\deg S
and 45\deg S.  In the northern hemisphere the 1998-2000 winds of
Hammel et al. (2001) seem slightly below the 2004 results, while the
2003 results of Hammel et al. (2005), are significantly more westward (more prograde)
than both 2000 and 2004 results, by $\sim$30 m/s between 35\deg N and
50\deg N.  Our results, which are dominated by our 2004 observations,
are shown as filled symbols in Fig.\ \ref{Fig:winddiff}. These are in
far better agreement with Voyager (open diamonds) and Karkoschka (1998)
(open circles) results than with the Hammel et al. (2001) and Hammel et
al. (2005) results.  Our results are inconsistent with a significant
sustained acceleration of wind speeds in the northern hemisphere.
Between 25\degx N and 30\degx N we do see a small average difference
from Karkoschka (1998) results, although the dispersion of the past
results in this region and the small statistical sample of new results
makes it unclear whether the difference is significant.

\begin{figure*}[!hbtp]\centering
\includegraphics[width=6in]{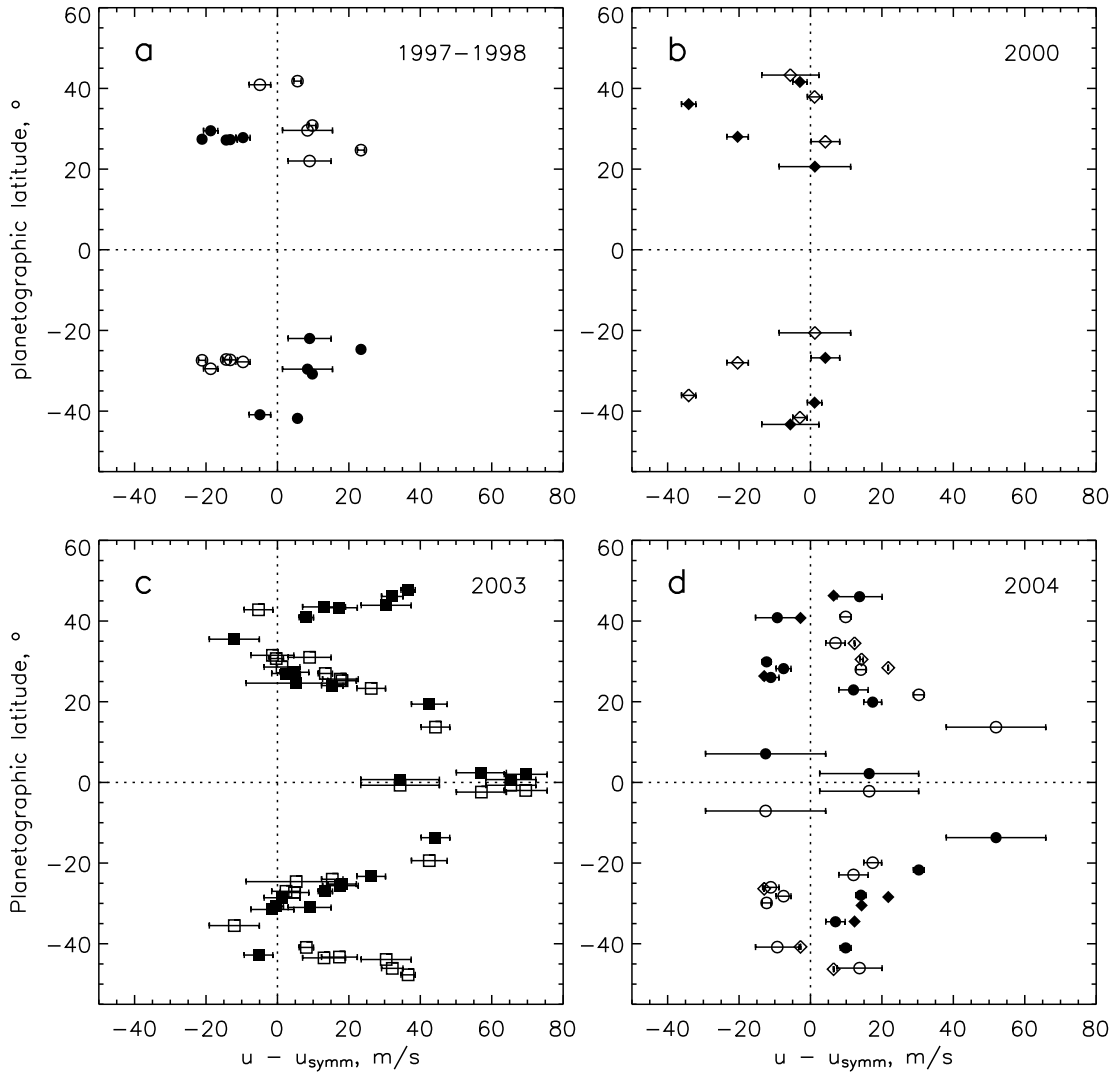}
\caption{Zonal wind speed relative to the symmetric fit displayed
in Fig.\ \ref{Fig:windbin} for different time periods: (a) 1997-98,
from Karkoschka (1998) and Hammel et al. (2001), (b) 2000, from
Hammel et al. (2001), (c) 2003, from Hammel et al. (2005), and (d)
2004, from this work, which includes highly accurate observations
from tracking clouds over a 1-month period (shown as diamonds).
Filled symbols are direct observations; open symbols are their reflections
about the equator.}
\label{Fig:asymm}
\end{figure*}

The temporal variations in asymmetry are clearly shown in Fig.\
\ref{Fig:asymm}, where we display four subgroups of wind observations
segregated by year of observation, using solid symbols for the direct
observations and open symbols for observations reflected about the
equator.  The four panels show symmetry characteristics for 1997-8,
2000, 2003, and 2004. Asymmetry about the equator is evident from the
separation in the wind direction between normal and reflected
observations. All the observations except 2003 results display a
significant asymmetry in which winds in the northern hemisphere are
less westward (less prograde) at latitudes between 20\deg and 40\degx, by $\sim$20 m/s.
The increase in northern hemisphere wind speeds reported by Hammel et
al. (2005) tends to make the zonal profile for that data set
relatively symmetric about the equator, as can be seen in Fig.\
\ref{Fig:asymm}c. The very slight asymmetry in these results is far less
 than found for observations before and after this period. Instead of
 a trend toward increasing symmetry as Uranus approaches equinox, our
 2004 results indicate a return to essentially the same degree of
 asymmetry that was observed in 2000 and 1997-8.

Some of the observed asymmetry may be due to asymmetries in altitude
of tracked features rather than to a real asymmetry in wind speeds.
Based on Voyager IRIS thermal observations of horizontal temperature
gradients, Flasar et al. (1987) infer vertical wind shears at mid
latitudes between 2.5 and 15 m/s per scale height. Since there are
approximately three scale heights between 3 bars ($\sim$37 km below
the 1-bar level) and 200 mb ($\sim$40 km above the 1-bar level), a
wind speed difference of 7.5-45 m/s might be expected between northern
and southern middle latitudes. This estimate for asymmetry based on
vertical wind shear covers the range that is actually observed,
presuming that southern cloud features are produced at the 3 bar level
and northern features are mainly near 200 mb, which is crudely
consistent with Paper II results.  However, the considerable noise in
the wind shear estimate, and uncertainty in its value deeper than 1
bar, prevent a definitive conclusion about the size of this
contribution to the observed asymmetry.  In addition, the mechanism
fails to explain why the asymmetry seems to be confined to mid
latitudes.  This might be related to latitudinal limitations to 
the respective roles of atmospheric and internal heat transport, as
discussed by Friedson and Ingersoll (1987).  As suggested by an anonymous
reviewer, some of the asymmetry might arise from differences in phase speeds
of cloud features related to differences in wave activity, both between
hemispheres and over seasonal time scales.  Phase speeds could easily
vary by 10's of m/s, depending on the nature of the wave mode that might be
involved.  Allison (1990) discusses this effect in the context of Jupiter's
near-equatorial atmosphere.

\section{Temporal changes in cloud characteristics}

\subsection{Temporal changes in morphology}

Analysis of Voyager observations by Rages et al. (1991) inferred the
existence of a southern polar cap of relatively thick clouds at about
the 1.2-1.3 bar level, increasing in optical depth from 0.7 at
22.5\degx S to 2.4 at 66.5\degx S. The bright cap was still prominent
in 1994, but declined significantly between 1994 and 2002.  This was
documented by Rages et al. (2004), who interpreted the changed
appearance as a decline in the optical depth of the methane cloud,
placed between 1.26 and 2 bars, which resulted in better views of
deeper cloud patterns at the 4-bar level. (Be aware that the abstract
of that paper erroneously states that the changes in cloud properties
occurred between 2 and 4 bars, where their model actually has no
cloud.)  In Paper II we show that the current near IR variations in
brightness with latitude cannot be explained by changes in cloud
properties in the 1.2-1.3 bar range, but must occur deeper in the
atmosphere, perhaps at the 4-bar level suggested by Rages et
al. (2004) as the source of the bright band features.  Between our
2003 and 2004 imaging observations little further change occurred in
the overall morphology of the cloud bands, but there were many changes
in discrete cloud features (Figs. \ref{Fig:2003images},
\ref{Fig:2004aimages}, and \ref{Fig:2004bimages}). Looking forward,
it is not reasonable to expect a permanent north south asymmetry to exist
in the Uranus' cloud structure.  The decline of the brightness of the
south polar cap makes the south closer to the north-south average I/F.
Likewise the relatively dark, recently exposed to view, northern hemisphere
might very well brighten.  If the contrast is due to solar forcing, it
is likely that the two hemispheres might be already near a maximum in contrast
and have started to move toward contrast reversal.

\subsection{Evolution of discrete features}

\begin{figure}[!hbtp]\centering
\includegraphics[width=3.5in]{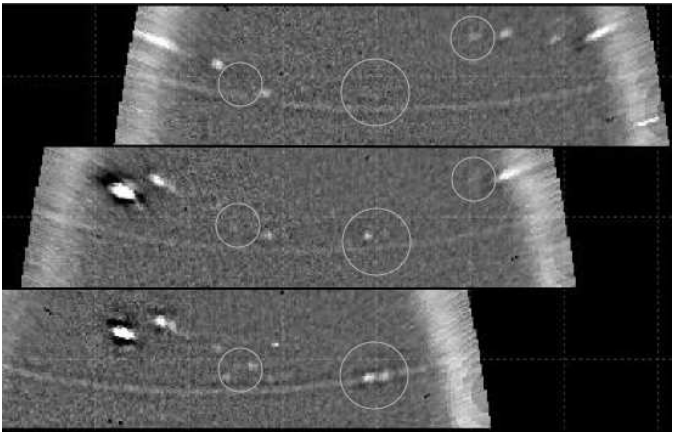}
\caption{Remapped 12 August 2004 H-band Keck images at 10:02 UT
  (bottom), 11:20 UT (middle), and 12:46 UT (top).  Each strip covers
  latitudes from 5\degx N to 35\degx N.  See text for description of
  evolving features within and near circular outlines.  The arcing
  feature is the Epsilon ring. The second and third strips are offset
  longitudinally to compensate for the zonal drift rate at 20\degx N.}
\label{Fig:rapid}
\end{figure}

Not all changes on Uranus proceed at the sedate pace one might expect
from the stability of cloud features observed by Karkoschka (1998). On
12 August 2004, we saw significant changes in the brightness of
discrete features within the span of an 0.3-1.5 hours.  Examples are
given in Fig.\ \ref{Fig:rapid}, which displays remapped image strips
centered on 20\degx N, at times 10:02 UT, 11:20 UT, and 12:46 UT.
Between the first and the second image, features within the left
circular outline essentially disappeared.  By the third image, no
trace of a cloud feature can be seen within the circle, while features
just outside the circle have brightened considerably. Since the
brightening and dimming features are of similar size, the changes
cannot be attributed to changes in seeing. A similar brightening can
be seen within the far right circular outline between the second and
third image.  Within the large circular outline near the center of the
figure there is another example of dramatic declines in cloud
brightness.  These rapidly evolving features share the characteristic
that they are relatively small, have relatively low brightness, and do
not reach high altitudes (they are not visible in K$'$ images).  

It is useful to consider what mechanisms might have time scales that
are compatible with these rapid changes.  Particles of 300$\mu$m
radius, inferred by Carlson et al. (1988) for the deeper clouds on
Uranus, have sedimentation times (time to fall a scale height) that
range from $\sim$3 hours at 1 bar to $\sim$10 hours at 7 bars.
However, particles moved below the condensation level of their parent
vapors by descending atmospheric flow will see a sudden decrease in
ambient relative humidity, resulting in relatively rapid evaporation.
Particles of 300-$\mu$m radius reaching the 60\% saturation level would
evaporate $\sim$10 minutes near 1.3 bars and in $\sim$30 minutes near
7 bars.  A vertically thin cloud layer can thus be changed drastically
by relatively small vertical motions.  A 1-km thick cloud at 1.2 bars,
would only need to move downward about 3 km to drop the relative
humidity to 80\%, at which point 300-$\mu$m particles could evaporate
within 20 min.  At 1 m/s, that 3-km drop would take less than an hour.
Once formed, clouds can also be dissipated by vertical and horizontal
shear, on time scales of the order of 1/($du/dy$) and 1/($du/dz$)
respectively.  The time scale for dissipation by horizontal shear has
a minimum of $\sim$13 hours, while time scale for dissipation by
vertical shear is 1/2 hour to 3 hours, based on the Flaser et
al. (1987) vertical wind shear estimates. Clearly, the key factor
controlling the lifetimes of the cloud features is the dynamics that
produces vertical motions.

The striking group of high-altitude northern
clouds appearing in the 12 July 2004 images (Fig.\
\ref{Fig:2004aimages}) provides an example of evolution in the
larger features, although over a longer time interval.  The
K$'$-bright cloud (136) at 26\degx N seems to be the origin of this
large complex of cloud features extending over $\sim$40\deg of
longitude, which amounts to a distance of $\sim$30,000 km.  There are
five major features associated with this complex including feature
136.  Just north of 136 is feature 137, which is at latitude 33.2\degx
N.  The measured drift rates for these features differ by
0.86$\pm$0.47\degx/h, and thus they do not travel together as fixed
pattern.  Among the other three components, 133 and 135 are both near
28.4\degx N and have slow drift rates between -0.34 and -0.48\degx/h.
The final component (134) is displaced by 0.6\deg to 27.8\degx N and
has a somewhat faster drift rate of -0.55$\pm$0.2\degx/h.

\begin{figure*}[!hbtp]\centering
\includegraphics[width=6.2in]{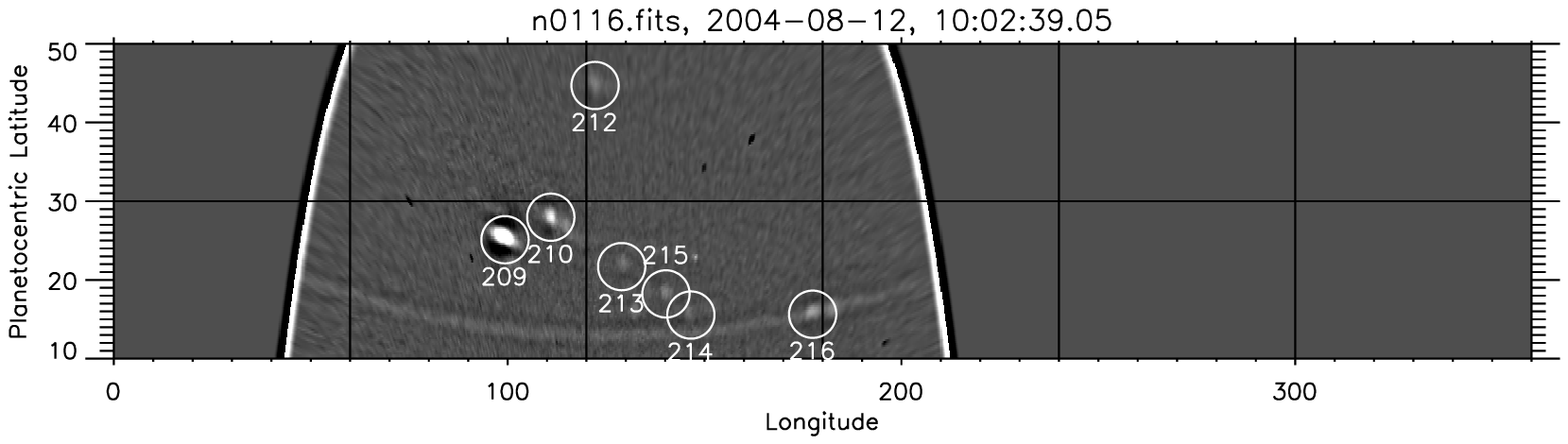}
\includegraphics[width=6.2in]{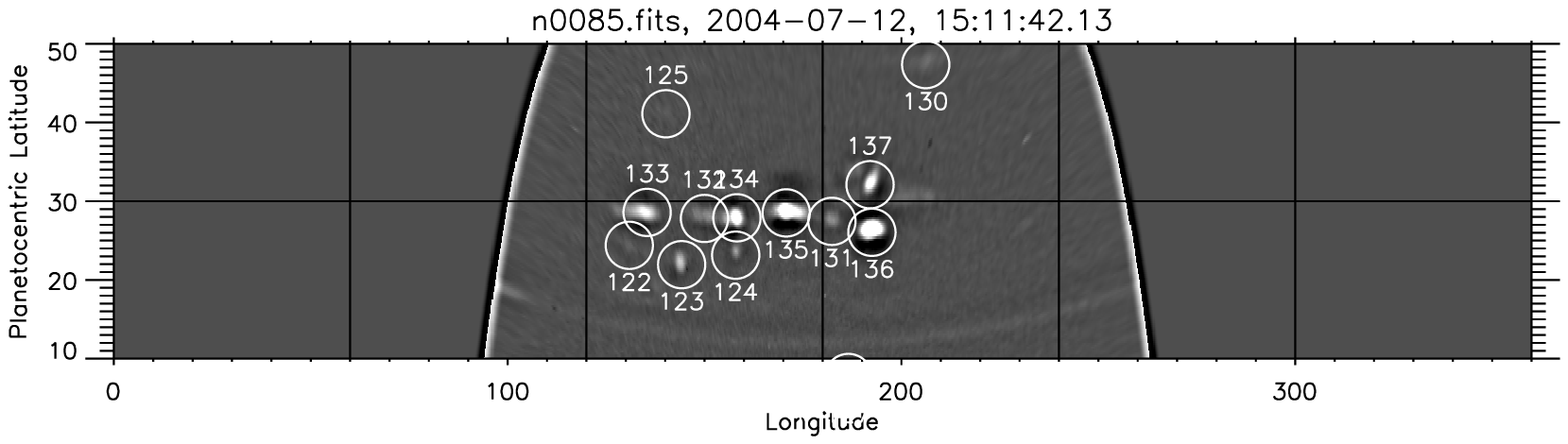}
\caption{Comparison of bright cloud features surrounding the 136/209
feature on 12 July (bottom) and 12 August (top) 2004.}
\label{Fig:136}
\end{figure*}

The fact that discrete feature 136 is by far the brightest cloud
feature in the K$'$ image suggests that it is also the most energetic
with respect to convective activity and may be the prime marker for a
substantial dynamical feature that is responsible for creation of all
these features.  Looking ahead to the August 11-12 observations, there
is another uniquely bright feature (209) which is at a similar
latitude (25.25\degx N).  That this is in fact the same feature
identified as 136 in the July images is further confirmed by the
continuity of the time varying longitude, assuming a constant drift
rate, which has a best fit value of -3.01$\pm$0.01\deg/day. Given the
relative small dispersion in drift rates of the various other features
associated with 136, one would expect to see them within 30\deg of
their original offsets from 136, after a month's delay, presuming they
lasted that long. As can be seen from Fig.\ \ref{Fig:136}, there are
few features within 30\deg of the original feature (labeled 209 in the
upper panel).

Feature 136 itself actually
declined significantly between 12 July and 12 August: in raw K$'$
images its peak contrast dropped by a factor of 4.6 and its
differential integrated brightness (Sromovsky et
al. 2000), which is far less seeing dependent, dropped by a factor of
2.2.  We estimated the size of this feature in deconvolved images.  In
both July and August it is about 5-6\deg in longitudinal extent, but
is so narrow in latitude that it can't be resolved. A line scan
through the feature has a full width at half maximum of $\sim$1.4\deg,
which corresponds to about 0.05$''$.  Thus opacity differences and
areal differences are both plausible contributors to the difference in
contrast.

\section{Characterization of a long-lived circulation feature near 34\degx S.}\label{Sec:char34}

\subsection{Morphology of S34}

\begin{figure*}[!hbtp]\centering
\includegraphics[width=6.4in]{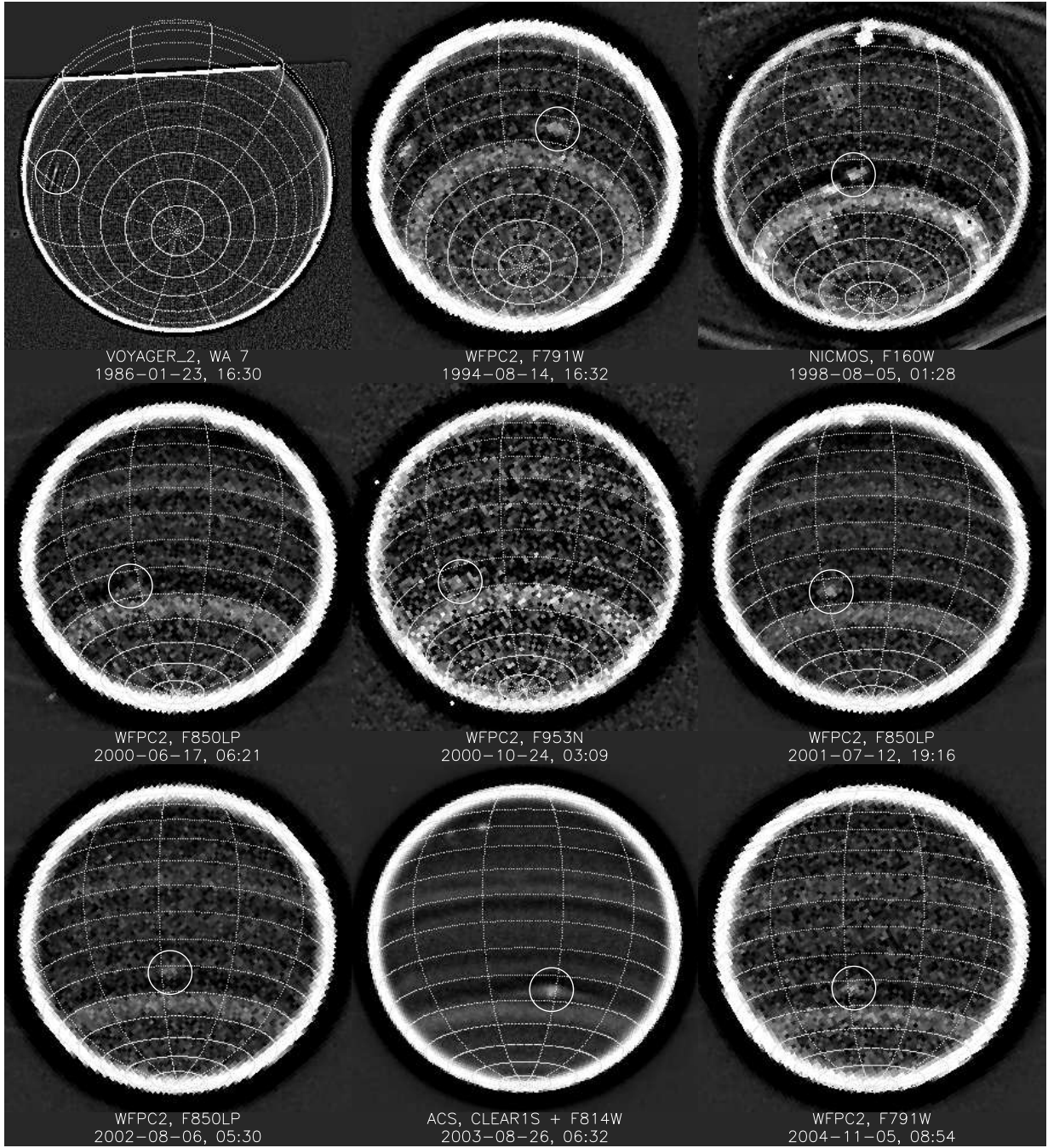}
\caption{S34 image samples from each of the periods
where it has been identified.  These are all high-pass filtered to
enhance contrast, which is quite low in Voyager Wide-angle Orange and
F953N WFPC2 images. See other figures for its appearance in recent
Keck images.}
\label{Fig:s34samples}
\end{figure*}

Among the targets listed in Table 6 and displayed in Figs.\
\ref{Fig:2003images}-\ref{Fig:2004bimages}, labels 301, 107, and 217 all
refer to the same circulation feature, even though it does not appear
at the same latitude during each time period. This is a significant
feature, first noted by Sromovsky and Fry (2004) to have a long
lifetime and unusual dynamics. For convenience we will hereafter refer
to it as S34, named for its mean latitude.  Target 301, the appearance
of S34 in our 2003 images, is not listed in Table 6 because it could
not be tracked long enough to yield an estimated wind speed error less
than 60 m/s.  However, its coordinates in 2003 images are listed in
Table 9, along with coordinates in prior images where we believe this
feature also appears.  The case is quite strong that S34 has been
present at least since the first detailed observations by Voyager in
1986, a time span of 18 years. Part of this case comes from continuity
of latitudinal and longitudinal motions, which are described in
subsequent sections. The other part comes from morphological evidence.
Observations with complete longitude coverage (in 1986, 2000, 2003,
2004) generally show only a single prominent feature between 30\degx S
and 40\degx S planetocentric latitudes.  The appearance of the feature
in HST and Voyager image samples from each of the time periods where
it can be located are provided in Fig.\
\ref{Fig:s34samples}.  According to Paper II, the vertical location of
this feature is generally mainly between 1.7 and 4 bars, and has a
contrast of 20-30\% in J and H bands.  Scattering by overlying haze
and methane cloud particles and Rayleigh scattering reduce contrast at
shorter wavelengths.

\begin{figure*}[!hbtp]\centering
\includegraphics[width=2.75in]{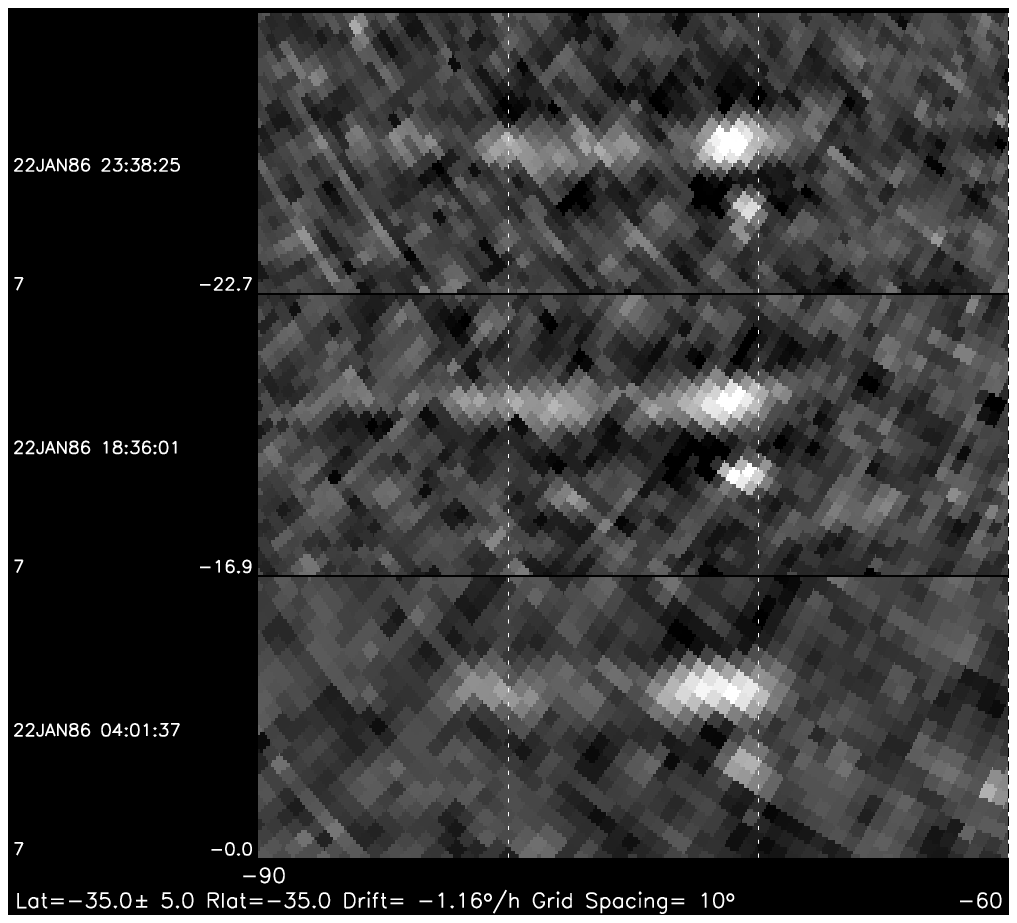}\hspace{0.2in}
\includegraphics[width=2.75in]{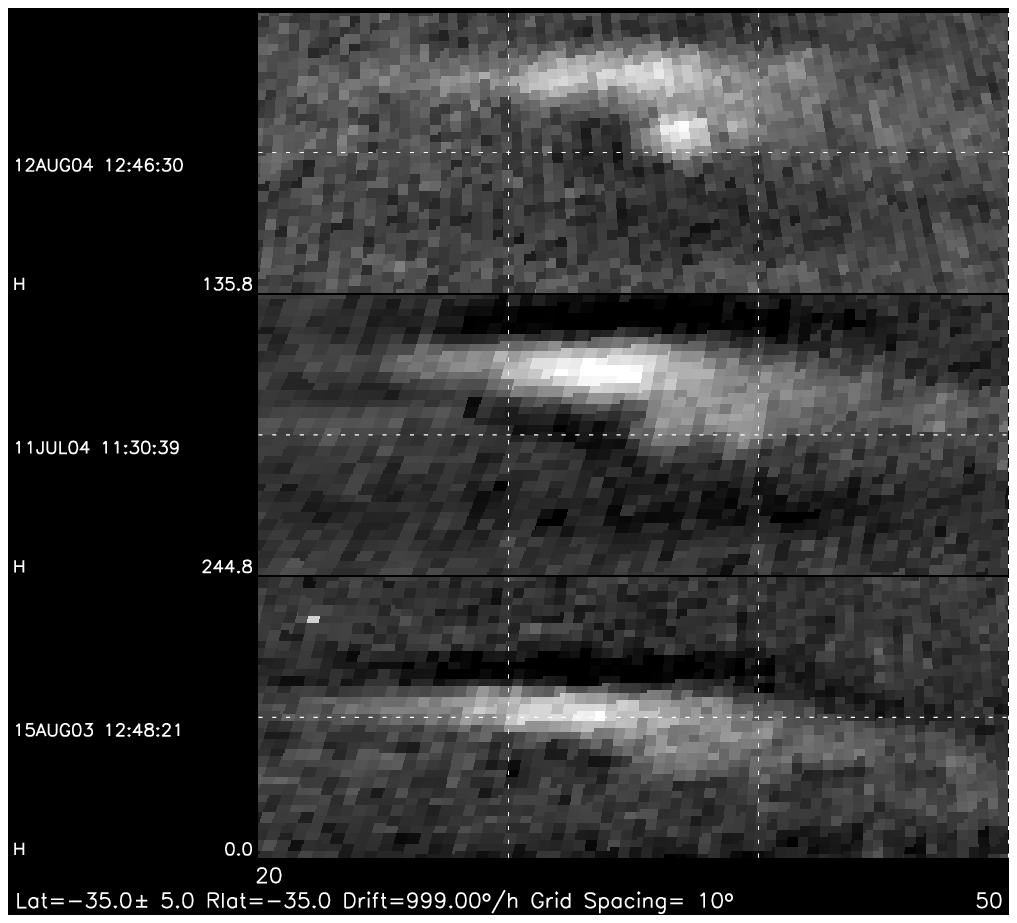}
\caption{Comparison of 34\degx S feature morphology in 1986 Voyager 2
 wide-angle orange filtered images (left), and 2004 H filter NIRC2
 Keck images. Both image groups are high-pass filtered to improve
 contrast. This is labeled as target 301, 107, and 217 in Figs.\
\ref{Fig:2003images}-\ref{Fig:2004bimages}.}
\label{Fig:morph34S}
\end{figure*}

Only recent Keck observations and Voyager 2 observations achieved
sufficient spatial resolution to capture detailed morphological
characteristics, which are compared in Fig.\
\ref{Fig:morph34S}. In both 1986 and in 2004 the feature is seen to have
two components, a long streaky northern component, which was thought
to be a plume by Voyager investigators (Smith et al., 1986) and which
extended over roughly 10-15\deg of longitude, and a small more
symmetric feature displaced 2.2-2.5\deg to the south of the main
feature. Intervening HST images by WFPC2 and NICMOS have detected
this feature but have not been able to resolve the detailed 2-component
morphology.  

\begin{table*}\centering
\caption{Observations of S34.}
\begin{tabular}{|c r c r c|}
\hline
              Time &  Relative  & Planetocentric          & Planetographic &\\
yyyy-mm-dd hh:mm:ss & Julian Day       & Latitude, \deg & \multicolumn{2}{c}{(East) Longitude,\deg \hspace{.5in} Source
\hspace{0.5in}   }\\
\hline
1986-01-22 04:01:34& -6745.312& -34.00$\pm$0.20&   288.60$\pm$0.5& Voyager WA Orange\\
1986-01-22 06:19:58& -6745.216& -34.30$\pm$0.20&   285.70$\pm$0.3& Voyager WA Orange\\
1986-01-22 07:25:34& -6745.170& -34.65$\pm$0.20&   284.80$\pm$0.3& Voyager WA Orange\\
1986-01-22 08:32:46& -6745.123& -34.70$\pm$0.20&   -76.75$\pm$0.3& Voyager WA Orange\\
1986-01-22 18:35:58& -6744.705& -33.80$\pm$0.20&   272.10$\pm$0.3& Voyager WA Orange\\
1986-01-22 19:28:46& -6744.668& -33.75$\pm$0.20&   271.10$\pm$0.3& Voyager WA Orange\\
1986-01-22 23:38:22& -6744.495& -34.80$\pm$0.20&   266.00$\pm$0.3& Voyager WA Orange\\
1986-01-22 00:21:34& -6744.465& -34.85$\pm$0.20&   264.65$\pm$0.3& Voyager WA Orange\\
1986-01-23 02:23:10& -6744.380& -34.80$\pm$0.20&   -97.30$\pm$0.3& Voyager WA Orange\\
1986-01-23 11:52:46& -6743.985& -33.60$\pm$0.20&   252.70$\pm$0.3& Voyager WA Orange\\
1986-01-23 14:02:22& -6743.895& -34.10$\pm$0.20&   249.50$\pm$0.3& Voyager WA Orange\\
1986-01-23 16:30:22& -6743.792& -34.25$\pm$0.20&   245.40$\pm$0.4& Voyager WA Orange\\
1994-08-14 13:20:16& -3618.924& -33.60$\pm$0.50&    56.60$\pm$2.0& HST WFPC2 F791W\\
1994-08-14 16:32:16& -3618.790& -34.30$\pm$0.50&    54.60$\pm$2.0& HST WFPC2 F791W\\
1998-08-05 01:28:40& -2167.418& -35.50$\pm$0.30&   206.00$\pm$1.0& HST NICMOS F160W\\
2000-06-16 15:52:14& -1485.818& -35.25$\pm$0.50&   105.30$\pm$2.0& HST WFPC2 F850LP\\
2000-06-17 06:21:14& -1485.215& -35.70$\pm$0.20&    85.80$\pm$2.0& HST WFPC2 F850LP\\
2000-06-17 19:13:14& -1484.679& -36.50$\pm$0.30&    71.00$\pm$3.0& HST WFPC2 F850LP\\
2000-06-18 00:03:14& -1484.477& -36.35$\pm$0.60&    67.30$\pm$2.0& HST WFPC2 F850LP\\
2000-06-18 12:34:38& -1483.955& -36.70$\pm$1.00&    49.00$\pm$2.0& s4 of de Pater et al. 2002\\
2000-10-24 03:09:13& -1356.348& -35.90$\pm$0.50&   218.10$\pm$3.0& HST WFPC2 F953N\\
2001-06-26 10:00:14& -1111.063& -34.90$\pm$0.50&    31.00$\pm$1.0& HST WFPC2 FQCH4N15\\
2001-07-12 04:49:14& -1095.279& -34.30$\pm$0.50&   -12.40$\pm$1.0& HST WFPC2 F850LP\\
2001-07-12 19:16:14& -1094.677& -35.00$\pm$0.30&   -27.40$\pm$1.0& HST WFPC2 F850LP\\
2001-07-13 13:02:14& -1093.936& -34.30$\pm$0.30&   -45.60$\pm$0.5& HST WFPC2 F850LP\\
2002-08-05 11:58:16&  -705.981& -32.00$\pm$0.30&    48.50$\pm$0.5& HST WFPC2 F850LP\\
2002-08-05 12:34:16&  -705.956& -32.40$\pm$0.40&    47.00$\pm$0.5& HST WFPC2 F850LP\\
2002-08-06 05:30:16&  -705.250& -32.50$\pm$0.40&    33.60$\pm$0.5& HST WFPC2 F850LP\\
2002-08-06 05:53:16&  -705.234& -32.40$\pm$0.40&    32.20$\pm$0.5& HST WFPC2 F850LP\\
2003-07-12 17:37:37&  -364.745& -35.60$\pm$0.30&    22.81$\pm$1.0& HST ACS F775W\\
2003-08-15 12:22:51&  -330.964& -35.90$\pm$0.30&    35.80$\pm$0.6& Keck NIRC2 H\\
2003-08-26 06:32:26&  -320.207& -35.50$\pm$0.30&    58.00$\pm$0.5& HST ACS F814W\\
2003-08-30 11:23:02&  -316.005& -36.30$\pm$0.50&   288.60$\pm$1.5& HST ACS F814W\\
2004-07-11 11:30:32&     0.000& -33.35$\pm$0.20&   280.59$\pm$1.0& Keck NIRC2 H\\
2004-07-12 15:27:37&     1.165& -33.40$\pm$0.20&   251.60$\pm$1.0& Keck NIRC2 H\\
2004-08-12 12:46:22&    32.053& -33.20$\pm$0.20&   171.90$\pm$0.6& Keck NIRC2 H\\
2004-11-05 08:54:16&   116.891& -33.00$\pm$0.40&   212.50$\pm$0.8& HST WFPC2 791W\\
\hline
\end{tabular}
\end{table*}

The origin of the 2-component  morphology is unclear,
although it does bear some resemblance to bright cloud features
associated with Great Dark Spots on Neptune (Sromovsky et al., 1993;
2002).  According to Stratman et al. (2001), bright clouds can be
generated by vortices in a manner analogous to the formation of
orographic clouds. These formations can be displaced from the center
of the vortex, and travel along with it, rather than following local
zonal wind profile.  Thus we speculate that there may be a vortex
circulation underpinning the two bright cloud elements that
characterize S34, both of which may be
orographic in nature.  The vortex would probably have a latitudinal
extent comparable to the latitude displacement between the two cloud
elements.  That would make it a little less than half the linear size
of the DS2 vortex on Neptune (Sromovsky et al., 1993), and only about
1/6 the size of Neptune's Great Dark Spot. It should be noted however that
no oval dark spot could be seen in any Voyager image of the Uranus 
feature.

\subsection{Motions of S34}\label{Sec:34move}

\begin{figure*}[!hbtp]\centering
\includegraphics[width=6.4in]{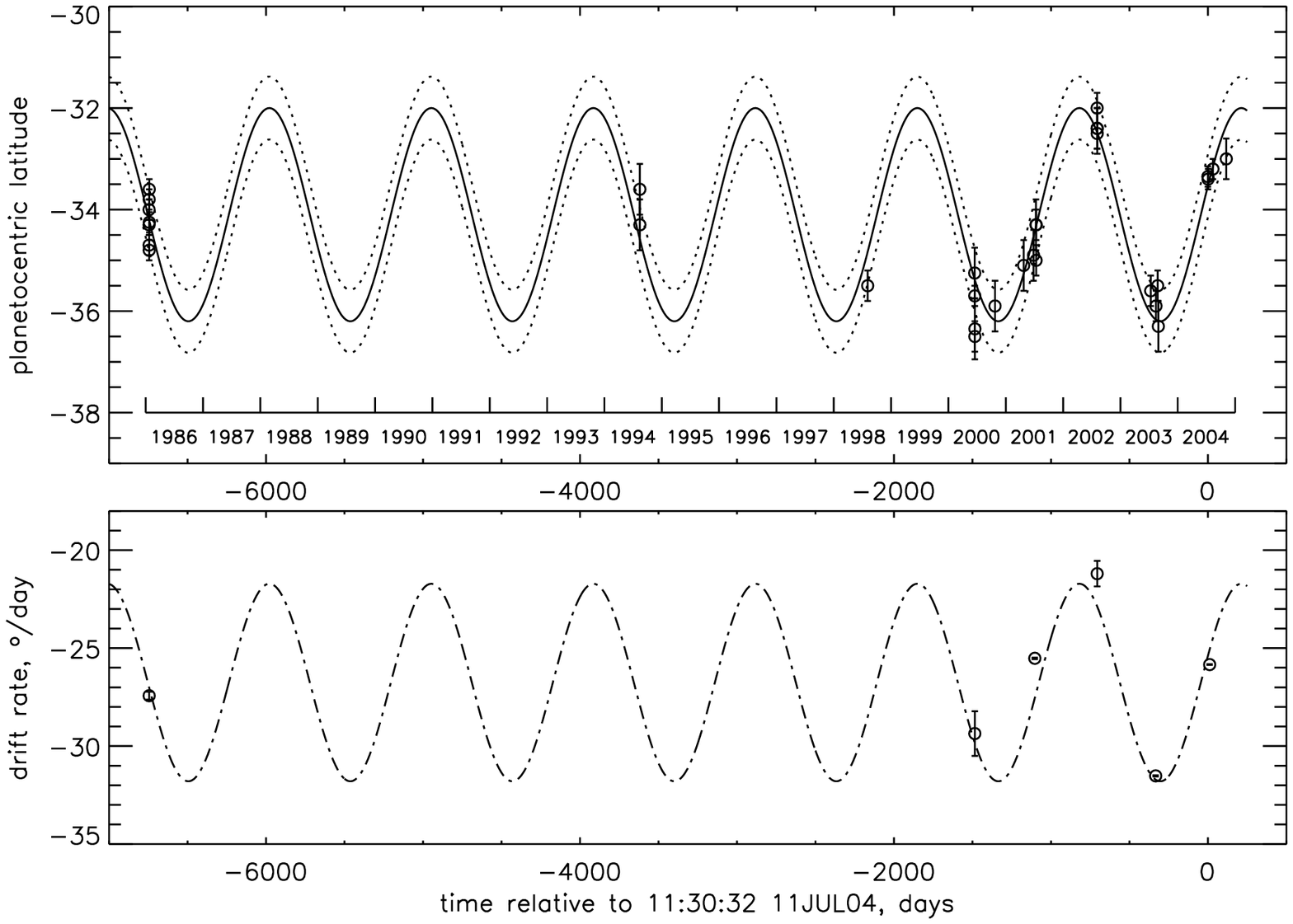}
\caption{Upper: 34S feature latitude vs time (symbols) and best-fit sinusoidal
model fit (solid line), with dotted lines indicating expected
variation due to inertial oscillations. The best-fit period is
1032 h and amplitude is 2.1\degx. Lower: 34S feature longitudinal drift
rate versus time for periods when accurate rates were determined (symbols).
The dashed curve is obtained from a fit of drift rate versus
latitude with input from the above latitude versus time fit.}
\label{Fig:34slowvar}
\end{figure*}

\begin{figure}[!hbtp]\centering
\includegraphics[width=3.5in]{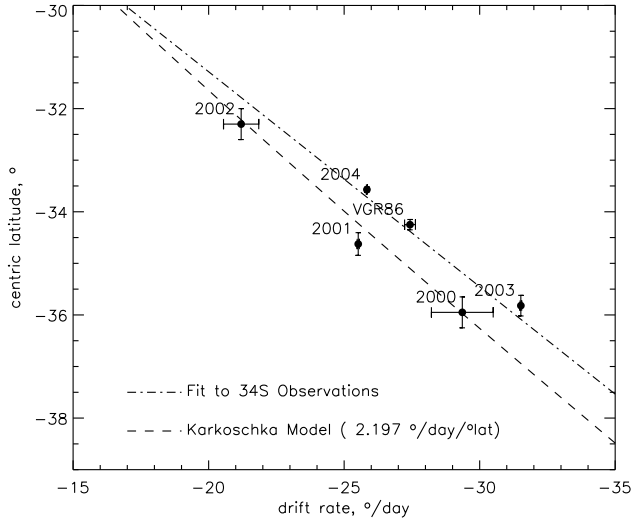}
\caption{Variation of eastward drift rate with latitude based on
  observations of the 34S feature.}
\label{Fig:latvsdrate}
\end{figure}

In recent years, it has become very clear that S34
does not travel at a fixed drift rate and does not remain at a fixed
latitude.  In fact, both vary by substantial amounts, as illustrated
in Fig.\ \ref{Fig:34slowvar}.  S34 positions are given in Table 9 and
S34 drift rates are given in Table 10 for time periods when observation
frequency has permitted accurate drift rate determinations. These are
the drift rates plotted in the lower panel of Fig.\
\ref{Fig:34slowvar}. The drift rate for 2003 was obtained by combining
observations in Keck and HST images taken by the Advanced Camera for
Surveys (ACS), acquired during observing program 9725 (E. Karkoschka,
PI).  At other times only a position observation was possible,
sometimes with a very inaccurate drift rate determination.  It is
roughly the case that the longitudinal drift rate is a simple function
of latitude that matches the zonal wind profile determined from tracking
the 34\degx S feature and features at surrounding latitudes.  This is illustrated in
Fig.\ \ref{Fig:latvsdrate}, compared to Karkoschka's empirical fit of the
zonal wind profile and to our fit to observations of S34.

Our empirical fits of the long-term variation in latitude ($\phi(t)$) and
drift rate ($d\lambda(t)/dt$) of S34 make two assumptions: (1) there is
a linear shear in the zonal wind and (2) the feature follows the zonal
wind variation with latitude.  The basic model equations that follow
from these assumptions are these:
\begin{eqnarray} \phi(t)&=& \phi_0 + c \sin(2\pi(t - t_0)/T)\label{Eq:lat}\\ 
\frac{d\lambda(t)}{dt}& =& d_{ref} + d_1 [\phi(t)-\phi_{ref}]\label{Eq:drate}
\end{eqnarray}  
where $T$ is the period of variation, $c$ is the latitudinal
amplitude, $\phi_0$ is the feature latitude at time $t_0$,
$\phi_{ref}$ is the latitude at which the zonal wind drift rate is
$d_{ref}$, and $d_1$ is the rate of change of drift rate with
latitude.  We used Eq.\ \ref{Eq:drate} to fit the drift rate results
assuming a reference latitude of $\phi_{ref}=-34.16$\deg and deriving
$d_{ref}$=-26.9$\pm$0.5\degx/day and $d_1=$2.4$\pm$0.5\degx/day/\degx
(lat.) by $\chi^2$ minimization. This fit is shown in Fig.\
\ref{Fig:latvsdrate} as a dot-dash line. We used Eq.\ \ref{Eq:lat} to
fit the slow latitude variation; this is shown as the solid curve in
the upper panel of Fig.\ \ref{Fig:34slowvar}.  The dotted curves
offset by $\pm$0.6\deg of latitude illustrate the bounds of a
short-period oscillation discussed below. The parameters of the
latitude fit that minimize $\chi^2$ are $\phi_0$=-34.1$\pm$0.1\degx,
$c=$2.1$\pm$0.2\degx, $t_0=$-45$\pm$10 days (for a time origin of
11:30:32 11 July 2004 UT), and $T=$1032$\pm$2 days. Plugging this
latitude function into the best-fit version of Eq.\ \ref{Eq:drate}
results in the dot-dash curve shown in the bottom panel of Fig.\
\ref{Fig:34slowvar}.

\begin{table*}\centering
\caption{Determinations of zonal drift rates for S34.}
\begin{tabular}{|c r c c c|}
\hline
             & relative     &                         &     longitudinal &\\
mean time   & time$^a$, days&   $<$latitude$>^b$,\deg &     drift rate$^c$, \degx/day&data source\\
\hline
1986-01-22 21:06:06& -6744.600& -34.25$\pm$0.10&   -27.43$\pm$0.20& Voyager WA Orange\\
2000-06-17 10:22:29& -1485.047& -35.95$\pm$0.30&   -29.36$\pm$1.14& HST WFPC2 F850LP\\
2001-07-04 14:38:14& -1102.870& -34.62$\pm$0.50&   -25.52$\pm$0.04& HST WFPC2 F850LP\\
2002-08-05 20:59:01&  -705.605& -32.30$\pm$0.30&   -21.20$\pm$0.65& HST WFPC2 F850LP\\
2003-08-13 11:58:58& -332.980 & -35.82$\pm$0.21&   -31.52$\pm$0.03& Keck + HST ACS\\
2004-07-22 13:14:50&    11.072& -33.32$\pm$0.50&   -25.87$\pm$0.02& Keck NIRC2 H\\
\hline
\end{tabular}
\parbox{5.in}{
\noindent$^a$Time is measured from 11:30:32 on 11 July 2004.\newline
$^b$Planetocentric latitude averaged about the mean time.\newline
$^c$Rate of change of east (planetographic) longitude.}
\end{table*}

\begin{figure}[!hbtp]\centering
\includegraphics[width=3.5in]{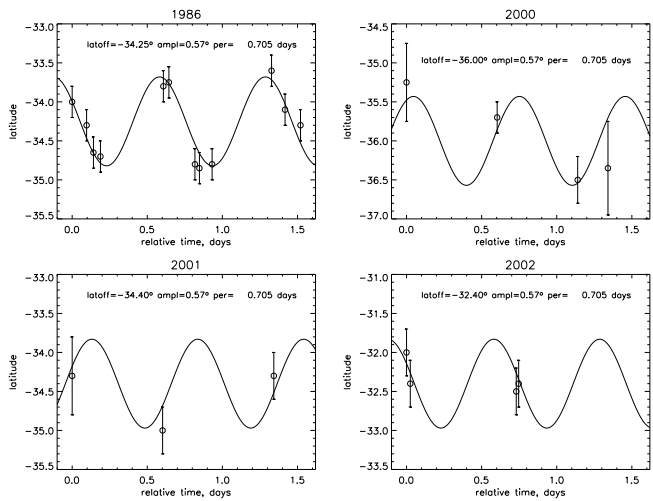}
\caption{Short term latitude variation of the 34\degx S feature:
  Voyager results (top) and HST WFPC2 results (2000-2002). The phases
  are adjusted independently for the four cases.}
\label{Fig:fastlatvar}
\end{figure}

The fit shown in Fig.\
\ref{Fig:34slowvar} illustrates that the oscillations
in latitude and drift rate are not absolutely uniform, and deviate
by more than the uncertainty in the measurements.
The existence of a real short-term variation in latitude is
confirmed by Voyager 2 observations, which provide the
high latitudinal accuracy and temporal sampling needed
to characterize it.  The 1986 position
measurements are plotted as a function of time in Fig.\
\ref{Fig:fastlatvar}. The empirical fit (solid line) is a simple sine 
wave of amplitude 0.57$\pm$0.05\deg and period 0.705$\pm$0.01 days
(16.92 hours).  Also shown in Fig.\ \ref{Fig:fastlatvar} is
that a sinusoidal variation of this magnitude and frequency
appears approximately consistent with the short-term variability seen in
2001, and possibly also with that seen in 2000 and 2002.  A
two-frequency oscillation is not unique to Uranus.  Sromovsky et
al. (2003) found a somewhat similar dynamical behavior in the 40\degx
S Neptune feature known as the Scooter.  However, for that feature the
long oscillation period (110 days) was ten times smaller and the
shorter oscillation period (22.5 days) was 30 times larger, and both
latitude amplitudes were quite small (0.16\deg and 0.315\deg
respectively).  Neptune's DS2 exhibited more substantial latitudinal
oscillations with an amplitude of 2.4\degx, which is quite close to
that of the 34\degx S feature on Uranus, though the 36-day period of
the DS2 is much shorter (Sromovsky et al. 2003).

Further constraints on the slow component of the 34\degx S oscillation
can be obtained using longitudinal position coordinates, which can
be useful even at times when temporal sampling does not permit
determination of an accurate drift rate.  The equation for the slow
variation of longitude follows directly from the previous expressions
for latitude and drift rate:
\begin{eqnarray}
\lambda(t)&=&\lambda_2 + d_2 \times (t-t_1) -
c_2\cos(\frac{2\pi}{T}(t-t_0)),\label{Eq:dlondt}
\end{eqnarray}  
where 
\begin{eqnarray}
\lambda_2&=&\lambda_1 + c_2\cos(\frac{2\pi}{T}(t_1-t_0)),\\
 d_2 &=& d_{ref}+d_1[\phi_0-\phi_{ref}],\label{Eq:d2}\\
 c_2 &=& \frac{d_1 c T}{2\pi},\label{Eq:c2}
\end{eqnarray}  
where $\lambda_1$ is the longitude at time $t_1$, where $c$, $d_1$, and
$\phi_{ref}$ are defined by Eqs. \ref{Eq:lat} and \ref{Eq:drate}. Inserting into Eqs.\
\ref{Eq:d2} and \ref{Eq:c2} the coefficient values derived from fitting
latitude vs time and drift rate vs latitude, we obtain the estimates
$d_2$ = -26.756$\pm$0.5\degx, and $c_2$ = 827$\pm$80\degx.  Since the
estimated longitudinal excursion relative to a uniformly
drifting longitude is two orders of magnitude greater than the
$\sim$1-2\deg accuracy with which feature longitude can be measured, these
observations can provide very tight constraints on the form of the
variation, provided temporal sampling is adequate to eliminate the
360\deg multiple ambiguity.  Unfortunately, for most time periods the
ambiguity is not easily resolved.  

\begin{figure*}[!hbtp]\centering
\includegraphics[width=6in]{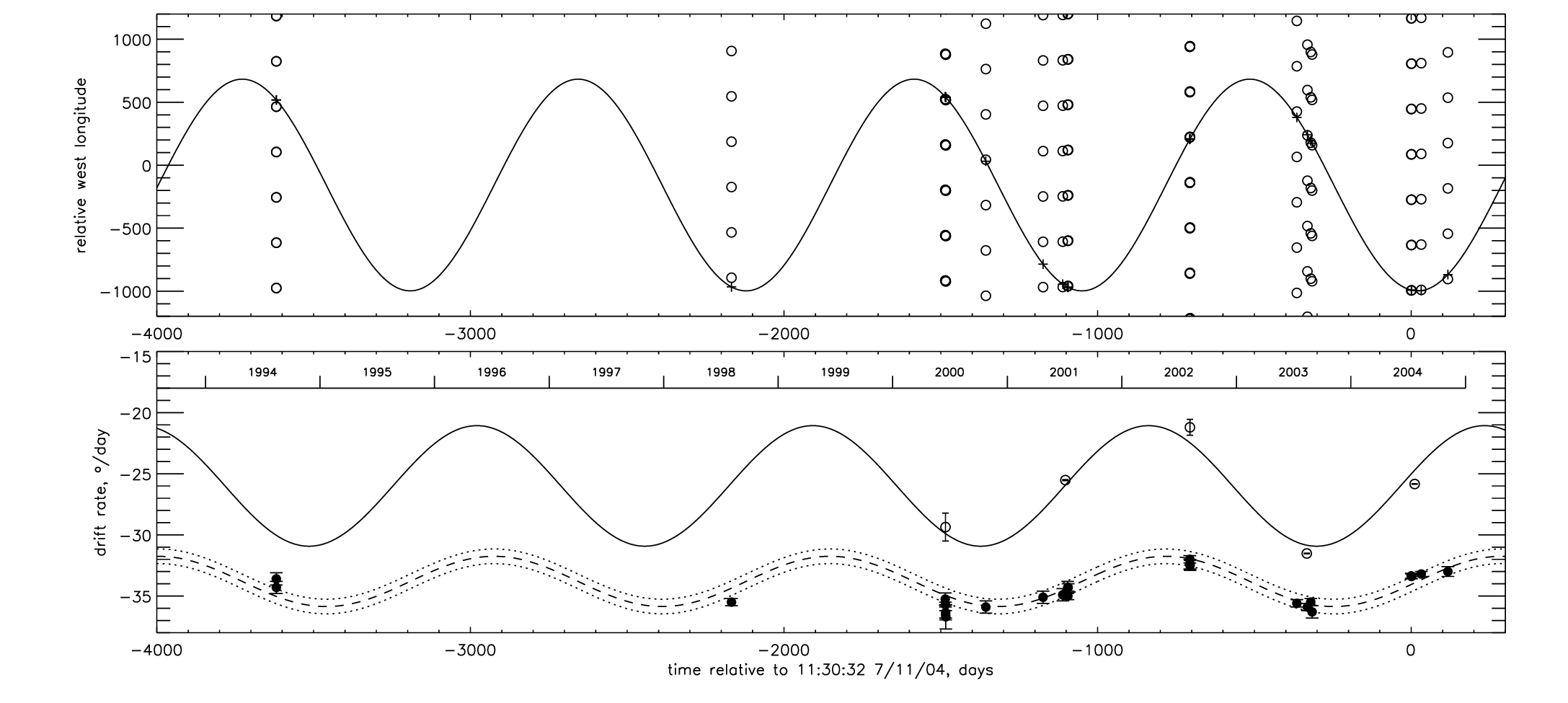}\par
\includegraphics[width=6in]{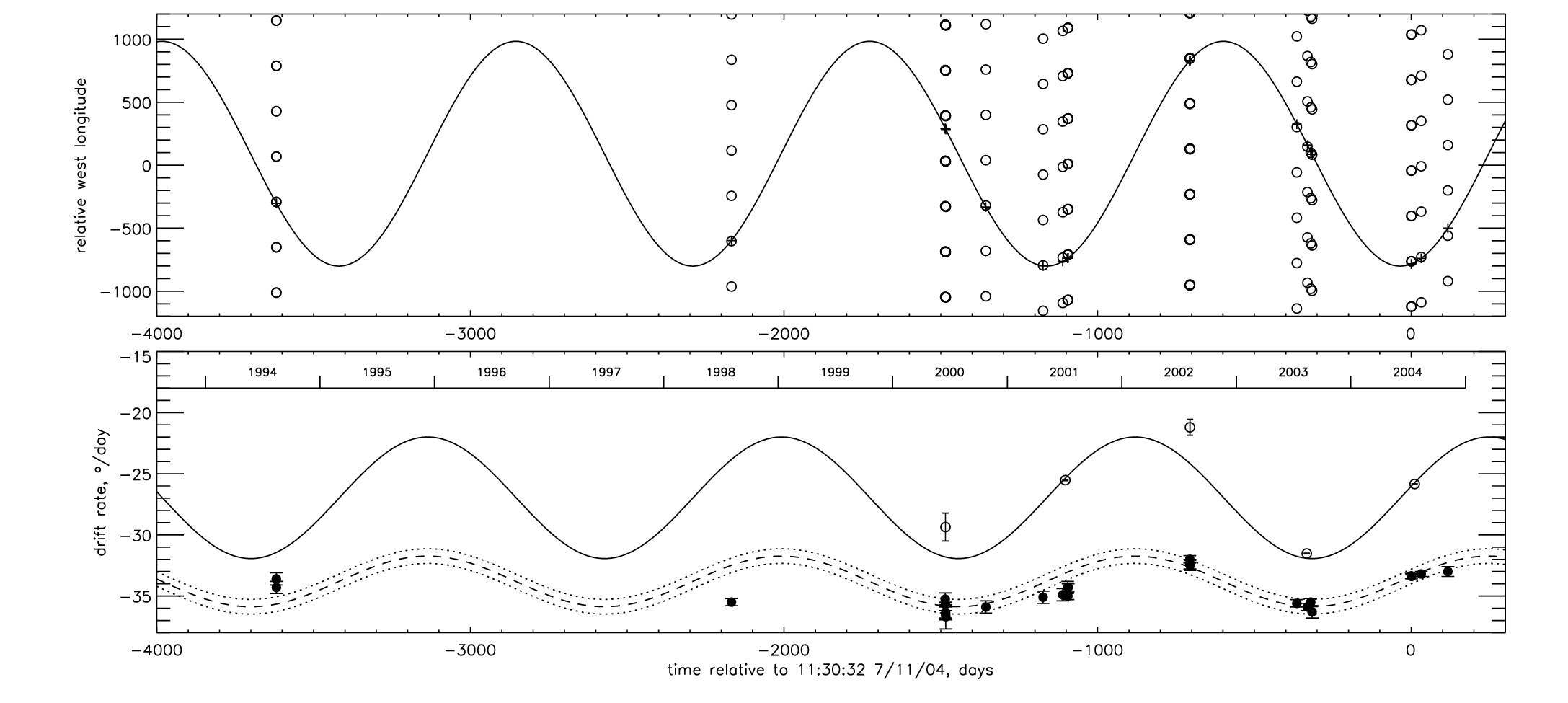}
\caption{Longitude of the 34\deg S feature vs time relative to fixed drift 
rates, and latitude and drift rates versus time for models with
periods of 1071 days (upper pair of figures) and 1128 (lower
pair). Longitude models are shown as solid curves. Fit coefficients
and $\chi^2$ calculations are given in Table 11. The latitude models
(dashed lines) are derived from the longitude fits using an inverted
form of Eq.\
\ref{Eq:dlondt}. For latitude
results the drift rate scale should be read as degrees centric. Dotted
curves surrounding the latitude fits show the range of the
short-period oscillation. Observations are plotted as
circles. Longitude uncertainties are much smaller than the symbol
size. }
\label{Fig:longfits}
\end{figure*}

Even allowing for the inherent longitude ambiguities, we were not able
make Eq.\ \ref{Eq:dlondt} accurately fit observations of both
longitude and latitude for the entire time period from 1986 to 2004
using a single period of sinusoidal variation.  Nor were the expected
coefficient values for $d_2$ and $c_2$ successful in providing more
than a very crude fit even for the 2000-2004 time period when sampling
is most dense.  Sample fits with altered coefficients are provided in
Fig.\ \ref{Fig:longfits}.  These fit coefficients are summarized in
Table 11, along with $\chi^2$ calculations for the more densely
sampled 2000-2004 period, and compared with expected values quoted
above. The $\chi^2$ computation includes latitude, longitude, and
drift rate differences. The latitude models are derived from the
longitude models
by inverting Eq.\ \ref{Eq:drate}.  The fit with the 1032-day
oscillation period provides the best match to latitudinal observations
over the entire time period but the worst overall fit quality over the
2000-2004 time period. The intermediate period fit ($T=$1071 d) provides a
relatively good match to longitudes, drift rates, and latitude
observations between 1994 and 2004, with the exception of the earliest
2001 point, which can only be explained if the feature is
misidentified.  This is certainly possible since the imaging during
this period covers only a small time interval that does not permit
examination of all longitudes.  The correct feature might very well
have been on the dark side of Uranus at this time.  The longer period
fit ($T$= 1128 d) in Fig.\ \ref{Fig:longfits} (and Table 11) is able
to fit this observation reasonably well, as well as most of the other
longitude observations, although several are well outside the
uncertainties of observation, especially the first 2000 point, which
is missed by nearly 100\degx.  From that point forward this fit
matches the most accurate drift rate observations quite well and does
fairly well matching the latitude observations.  At earlier times,
however, latitude observations are quite inconsistent with this fit.

\begin{table*}\centering
\caption{Model fits to slow variations in east longitude of S34.}
\begin{tabular}{|c c c c c c c c c|}
\hline
osc.    &                  & osc.      & osc.          & linear        & long.       & &&\\
period  & $<$drift rate$>$ & ampl. & time ref.$^a$ & time ref.$^a$ & offset   &  \multicolumn{2}{c}{Fit quality} &\\
$T$ (days)& $d_2$ (\degx/day)      & $c_2$ (\deg)  & $t_0$ (days) & $t_1$ (days) & $\lambda_2$ (\degx) &
 $(\chi^2)^b$& $(\chi^2)^c$& \\
\hline
1032   & -27.044$^d$ & 827   & -45   & -1094.7$^d$ &  15.0$^d$  & 55,826 &&\\
1071   & -25.996 & 840   & +21   & -1356.6 &  22.7  & 16,372 & 9,398$^c$&\\
1128   & -26.975 & 892   & -35   & -1356.4 & 271.3  & 20,948 &&\\
\hline
\end{tabular}
\par
\parbox{4.55in}{
\noindent$^a$Time is measured from 11:30:32 on 11 July 2004.\newline
$^b$Computed for 2000-2004 combining longitude, latitude, and drift-rate differences.\newline
$^c$This value is for 2000-2004 but excludes the earliest 2001 observation.\newline
$^d$Adjusted for best fit; other parameters in this row are from the long-term latitude fit.}
\end{table*}

Even the best fit in Table 11 has a $\chi^2$ dominated by longitude
differences, which have RMS values that are 20 times the expected
errors.  The differences between observations and these simple fitting
functions are found to be more systematic than random.  For example,
the $\chi^2$ value for the longer period fit can be halved by adding a
second modulation with a 740-day period and an 80\degx amplitude.  It
is also possible to get much more accurate fits by fitting smaller
time spans.

The inescapable conclusion from attempts to fit longitude observations
is that the motion cannot be accurately describe by a simple
sinusoidal variation in longitude about a constant drift rate.  It may
be that the motion is sinusoidal with a slowly varying frequency, or
that there are multiple frequencies present. A useful investigation of
these issues is limited by large temporal gaps that introduce
rotational ambiguities.  Additional observations of S34 positions at a
frequency of several times per year for the next five years or more
would provide excellent constraints on the nature of this variation.

\subsection{Interpretations of short-period oscillations of S34}

The short-period oscillation of S34 has characteristics
reasonably consistent with an inertial oscillation, which is a
circular flow in which there is a balance between the Coriolis
force and the centrifugal force, and other horizontal forces are 
negligible. The period of an inertial oscillation is given by \begin{equation}
 T_{\mathrm{in}} = T_{rot} / (2 \sin \phi_{pg}), \label{Eq:tinert}
\end{equation}
where $T_{rot}$ is the rotational period at latitude $\phi_{pg}$
(Holton, 1972). This evaluates to 15.84 h (0.66 earth days) at a centric
latitude of -34\degx, which is within 1.08 h of the
best-fit period.  The radius of curvature of an inertial oscillation
is proportional to the speed of the flow and inversely proportional to
the Coriolis parameter ($2\Omega \sin \phi_{\mathrm{pg}})$.  Measured
in angle instead of physical distance, the oscillation amplitude in
longitude is thus expected to be larger by the ratio of the
latitudinal radius of curvature to the radius of rotation at the
latitude of interest, which evaluates to $1/\cos\phi$ for a spherical
planet.

\begin{figure*}[!hbtp]\centering
\includegraphics[width=4.5in]{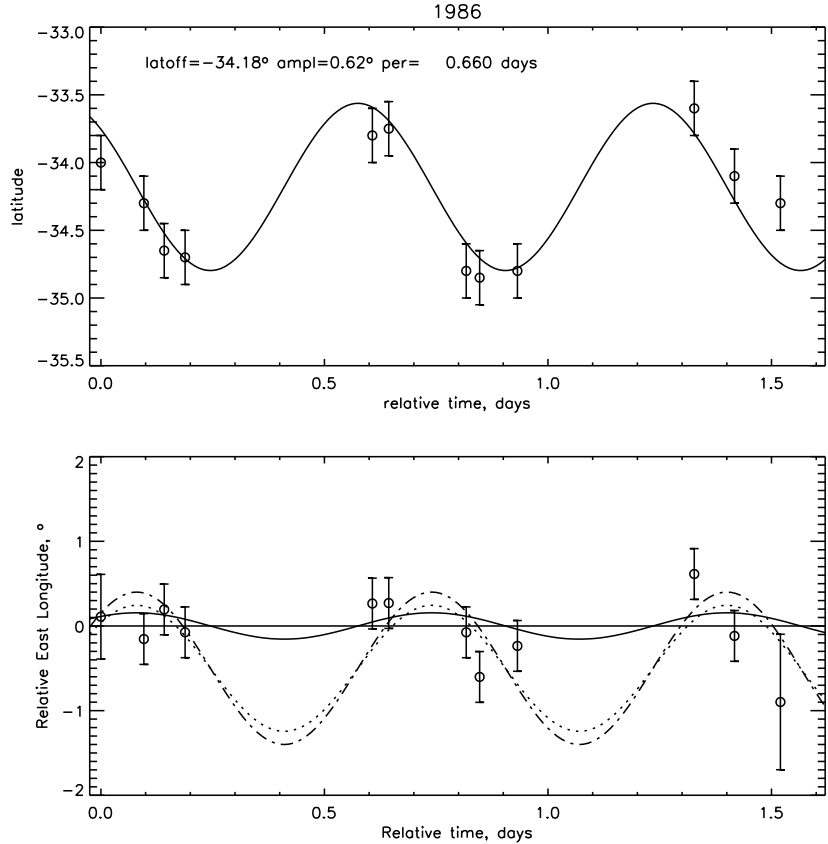}
\caption{Short term latitude and longitude variation of the 34\degx S
  feature based on Voyager results (top) and longitude variation
  relative to the best fit uniform drift rate. The solid line is the
  variation expected from the latitude shear only, the dotted line is
  the motion expected without latitudinal shear, and the dot-dash line
  is the sum of the two components.}
\label{Fig:inertial}
\end{figure*}

The longitude observations are somewhat too uncertain and too poorly
distributed to provide much additional independent confirmation of the
circular nature of the oscillation, although the results are
reasonably consistent with such an oscillation. This is shown in Fig.\
\ref{Fig:inertial}, where latitudinal and longitudinal
deviations are compared to model variations that are constrained to match the
inertial period of 0.66 days, rather than the best-fit longitudinal
period of 0.705 days.  The longitudinal variations are deviations from
a best fit constant drift rate computed from the 1986 Voyager
observations.  The solid line in the lower panel is the variation
expected if the longitudinal motion simply followed the latitudinal
shear in the zonal wind.  The dotted curve is the variation expected
from the inertial oscillation alone, in the absence of meridional wind
shear.  The dot-dash curve is the sum of the two contributions.  The
temporal sampling of the observations does not provide a very good
constraint on the longitudinal model, but does seem to favor the
shorter period computed for the ideal inertial oscillation, while the
latitudinal variations clearly prefer a slightly longer period.  The
peak longitude variation about the mean drift longitude is also more
consistent with the inertial model expectation, although the
uncertainties are too large to make this compelling evidence that this
is the dominant mechanism.

For an inertial oscillation, the instantaneous drift rate would be expected
to vary by a significant amount when measured over short time intervals.  The
relative variation is given by the relation:\begin{equation}
  (d\lambda/dt)_{\mathrm{in}} =  \frac{2\pi c_{\mathrm{in}}}{T_{\mathrm{in}}\sin\phi_{pg}}
      \sin(\frac{2\pi}{T_{\mathrm{in}}}t)
\end{equation}
where $c_{\mathrm{in}}$ is the latitudinal amplitude of the inertial
oscillation and $t$ is time measured from a zero of the latitudinal
oscillation at which latitude is increasing.  For $c_{\mathrm{in}}$ =
0.6\degx, $\phi$=-34\deg ($\phi_{pg}$=-35.24\degx), and $T_{\mathrm{in}}$=
0.66 days, the amplitude of the drift rate modulation evaluates to
9.9\degx/day.  This is somewhat surprising given that the amplitude of
longitude modulation is only $\sim$1\degx. If this is sampled at positive
and negative peaks, which are separated by only 0.33 days, it represents
a drift rate deviation of $\sim$2\degx/0.33 days = 6\degx/day. When these
deviations are sampled over a period of 0.99 days, the drift rate perturbation
is reduced to a maximum of $\sim$2\degx/day.  Since both negative and positive
samples would be expected over two successive nights of observation, a significantly
smaller perturbation would generally be expected. Nevertheless, some of the
deviations in drift rate from the model fit shown in Fig.\ \ref{Fig:34slowvar} could
be due to imperfectly averaged inertial oscillations.

\subsection{Interpretation of slow oscillations of S34}

\begin{figure}[!hbtp]\centering
\includegraphics[width=3.4in]{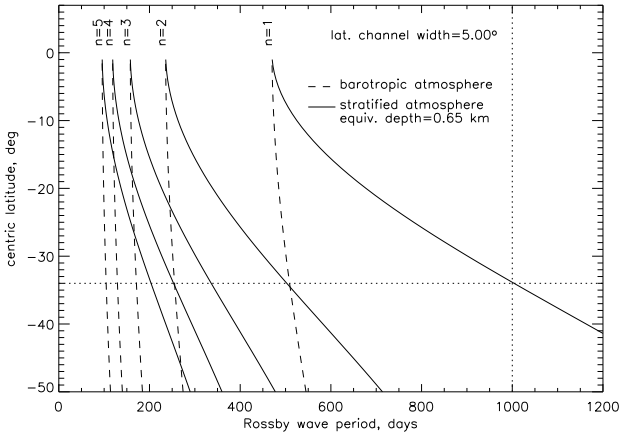}
\caption{Rossby wave periods for a wave confined to a 5\degx-wide
  latitude band as a function of zonal wave number (per
  circumference), for barotropic (dashed) conditions and for
  stratified conditions (solid) for an equivalent depth of 0.65 km.
  The observed period near 1000 days is best matched by the n=1
  non-barotropic option.}
\label{Fig:rossby}
\end{figure}

Among well studied transverse atmospheric wave motions, only Rossby waves can
yield long enough periods to match S34 observations.  The
Rossby wave arises from a gradient in planetary vorticity.  It is a
transverse wave that produces flow curvature that increases and
decreases its relative vorticity to compensate for changes in
planetary vorticity that result from deviations in latitude.  This is
the wave mode that is evident in the meandering jet stream on earth.
In the northern hemisphere on earth, the phase speed is westward
relative to the zonal flow.  The period of the latitudinal oscillation
is given by:
\begin{equation} T=
\frac{2\pi (k^2 + m^2 +f^2/gh)}{k \beta},
\label{Eq:rossper}
\end{equation}
where the $k$ and $m$ are the zonal and meridional wavenumbers
($2\pi/$wavelength), $\beta=2\Omega \cos(\phi_{\mathrm{pg}})/a$ is the
derivative of the Coriolis parameter with respect to latitude, $a$ is
the local radius of curvature in the latitudinal direction, 
$g$ is the acceleration of gravity, and $h$ is an effective thickness
parameter related to vertical stratification and involvement of
vertical motions with the horizontal wave motions.  This form, which ignores
the small curvature of the zonal wind profile relative to $\beta$, is
consistent with Pedlosky (1979).
The last term in the numerator can be written in a
number of ways, all dependent on the degree to which vertical motions
participate in the wave motion. Here we use the expression that is the
inverse square of the Rossby deformation radius.  Two cases are
illustrated in Fig.\
\ref{Fig:rossby}, both of which assume a meridional half-wavelength of
5\degx.
The first case is pure barotropic, which has a maximum period of only
about half what is observed for S34, even for a zonal
wavenumber of unity (per circumference). The second case includes the
vertical motion term, with the effective depth adjusted to produce a
period of 1000 days at 34\degx S for the longest of the possible
planetary wavelengths.  The effective depth, which is only 0.65 km,
cannot be interpreted as the true physical depth of the motion (see
Pedlosky 1979).  From this comparison, it is clear that a pure
barotropic Rossby wave cannot be responsible for the long-period
motion of S34, although a Rossby wave with coupled
vertical motions is conceivable.

\section{Conclusions}

Near-infrared adaptive optics imaging of Uranus by the Keck 2
telescope during 2003 and 2004 has revealed unprecedented numbers of
discrete cloud features, which is partly due to improved adaptive
optics performance and partly due to more of Uranus' northern
hemisphere becoming visible as the planet approaches its 2007 equinox.
Increased cloud development might also play a role.
Specific results of our investigation are based primarily on Keck
observations we made in 2003 and 2004, but also on HST and Voyager
observations.  Our main conclusions are as follows:

\begin{enumerate}

\item Recent near IR adaptive optics imaging from the Keck telescope
achieved angular resolutions near 0.05$''$, facilitating the detection
of numerous discrete cloud features, which display a strong latitudinal
variation in distribution and cloud characteristics, in reasonable accord
with recent prior observations from the Keck telescope (de Pater et al. 2002,
Hammel et al. 2005). The northern mid latitude features reach the
highest altitudes and show the greatest K$'$ contrast and sharpest
horizontal structure, while equatorial
features are dim and fuzzy. Features in the southern hemisphere
were fewer in number, and none were seen south of a bright band near
45\degx S.

\item Measurement of cloud motions with estimated errors less than
50 m/s were obtained for 50 northern hemisphere
features and 20 southern features.  This has extended
prior circulation measurements up to 60\degx N, where we found a
maximum wind speed of 240$\pm$50 m/s.  Prior results were limited to
46.4\deg N (Hammel et al. 2005).  We confirmed the presence of an
asymmetry in the zonal circulation of Uranus, originally pointed out
by Karkoschka (1998), and improved its characterization.  We find that
the asymmetry is mainly evident at mid latitudes where winds are more
westward (prograde) in the southern hemisphere than in the northern hemisphere,
with an average difference of $\sim$20 m/s.

\item There is no clear indication of sustained long-term change in
wind speed between 1986 and 2004, although results of Hammel et
al. (2001) based on 2001 HST and Keck
observations are $\sim$10m/s less westward (less prograde) between 30\degx S and
45\degx S than earlier and later
results, and results of Hammel et al. (2005) based on 2003 Keck observations
are $\sim$30 m/s more westward (more prograde) between 20\degx N and
50\degx N than both 2000 and our 2004 results.  

\item We found a large dispersion in lifetimes of discrete features
that contrasts with Karkoschka's characterization of features he observed with
the HST NICMOS camera as relatively unchanged over a 100-day period
(Karkoschka 1998). Some of the discrete cloud features we observed
have relatively short lifetimes of $\sim$1 hour, some have disappeared
within a month, while others have lasted at least one month, and S34
seems to have persisted for nearly two decades. The shortest
lifetimes were observed for smaller low-contrast
features that would not be visible in NICMOS images.

\item Based on detailed morphological similarities
between high-resolution Voyager images in 1986 and Keck images in
2004, based on the continuity of latitudinal excursions as determined from
HST and Keck imaging in 1994, 1998, 2000, 2001, 2002, 2003, and 2004,
and from the fact that only a single feature of consistent morphology
has appeared within the latitude band within which S34 is restricted,
we believe that S34 has existed at least since 1986, which, if true,
would make it the longest lived discrete feature on Uranus.

\item S34 oscillates in latitude between 32\degx S and 36.5\degx S, with a
period of $\sim$1000 days, which may be a result of a non-barotropic
Rossby wave.  Its longitudinal drift rate varied between
-20\degx E/day and -31\degx E/day in approximate accord with the
latitudinal gradient in the zonal wind profile, exhibiting behavior
similar to that of the DS2 feature observed on Neptune (Sromovsky et
al., 1993).
The Uranus feature also exhibits a superimposed much more rapid
oscillation, with a latitudinal amplitude of 0.57\deg and period 0.7
days, which is reasonably consistent with an inertial oscillation.
The inertial oscillation has a longitudinal amplitude of $\sim$1\degx, which
leads to a large variability in longitudinal drift rates measured over short time
periods.

\end{enumerate}

\section*{Acknowledgments}

This research was supported by NASA Planetary Astronomy Grant
NAG5-12206. We are grateful to the Keck AO staff, to Support
Astronomers Randy Campbell, Grant Hill, and Robert Goodrich, Observing
Assistants Chuck Sorenson, Steven Magee, and Terry Stickel,
and to the W. M. Keck Observatory, which is made possible by the generous
financial support of the W. M. Keck Foundation.   We thank those of
Hawaiian ancestry on whose sacred mountain we are privileged to be
guests. Without their generous hospitality none of our groundbased
observations would have been possible. This research was partly based
on archived Hubble Space Telescope observations, obtained at the Space
Telescope Science Institute, which is operated by the Association of
Universities for Research in Astronomy, Inc., under NASA Contract
NAS5-26555. We also wish to acknowledge the USGS, its Astrogeology
Team, and other contributors for their development and support of the
Integrated Software for Imagers and Spectrometers (ISIS) system, which
we used in the analysis of Voyager 2 images.

\section*{References}

\begin{description}
\itemsep=0pt
\item[] Acton, C. H. 1996. Ancillary data services of NASA's Navigation and Ancillary
Information Facility. Planet. Space Sci., 44, 65-70.

\item[] Allison, M. 1990. Planetary Waves in Jupiter's Equatorial
Atmosphere, Icarus 83, 282-307.

\item[] Allison, M., R.\ F. Beebe, B.\ J. Conrath, D.\ P. Hinson, A.\ P. Ingersoll 1991.
Uranus atmospheric dynamics and circulation. In Uranus (J.\ T. Bergstralh, E.\ D. Miner,
and M.\ S. Matthews, Eds.), pp 253-295, Univ. of Arizona Press, Tucson.

\item[] Baines, K.\ H., M.\ E. Mickelson, L.\ E. Larson, and D.\ W. Furguson 1995.
The abundances of methane and ortho/para hydrogen in Uranus and Neptune: Implications
for new laboratory 4-0 H$_2$ quadrupole line parameters. {\it Icarus \bf 114}, 328-340.

\item[] Carlson, B.\ E., W.\ B. Rossow, and G.\ S. Orton  1988. Cloud Microphysics of
the Giant Planets, J. Atmos. Sci. 45, 2066-2081.

\item[] Danielson, G.E, Kupferman, P.N., Johnson, T.V., Soderblom, L.A, 1981. Radiometric
performance of the Voyager cameras. J. Geophys. Res. 86, 8683-8689.

\item[] de Pater, I., S.\ G. Gibbard, B.\ A. Macintosh, H.\ G. Roe, D.\ T. Gavel,
and C.\ E. Max  2002. Keck Adaptive Optics Images of Uranus and Its Rings. Icarus
160, 359-374.

\item[] Flasar, F.\ M., B.\ J. Conrath, P.\ J. Gierasch, and J.\ A. Pirraglia
1987. Voyager Infrared Observations of Uranus' Atmosphere: Thermal Structure
and Dynamics. J. Geophs. Res. 92, 15011-15018.

\item[] Forsythe, K.J., M.S. Marley, H.B. Hammel, E. Karkoschka 1999.
The Changing Face of Uranus, {\it Bull Am. Astron. Soc. \bf 31}, 1153.

\item[] Friedson, J., and A.\ P. Ingersoll 1987. Seasonal Meridional Energy Balance and
Thermal Structure of the Atmosphere of Uranus: A Radiative-Convective-Dynamical Model.
Icarus 69, 135-156.

\item[] Hammel, H.\ B., K. Rages, G.\ W. Lockwood, E. Karkoschka, and I. de Pater
2001. New Measurements of the Winds of Uranus. Icarus 153, 229-235.


\item[] Hammel, H.\ B., I. de Pater, S. Gibbard, G.\ W. Lockwood, and K. Rages
2005.  Uranus in 2003: Zonal Winds, Banded Structure, and Discrete Features.
Icarus 175, 534-545. 

\item[] Hanel, R.\ A., B.\ J. Conrath, F.\ M. Flasar, V. Kunde, W. Maguire,
J. Pearl, J. Pirraglia, R. Samuelson, D. Cruikshank, D. Gautier, P. Gierasch, L. Horn, 
and P. Schulte 1986. Infrared observations of the Uranian system. Science 233, 70-74.

\item[] Holton, J.\ R. 1972. An Introduction to Dynamic Meteorology. Academic Press,
New York, 319 pages.

\item[] Karkoschka, E. 1998.  Clouds of High Contrast on Uranus.
{\it Science \bf 111}, 570-572.

\item[] Karkoschka, E. and M.G Tomasko 1998.  Bright Discrete and Zonal Features on Uranus.
 {\it Bull. Am. Astron. Soc. \bf 30}, 1056-1057.

\item[] Limaye, S.\ S., and L.\ A. Sromovsky 1991. Winds of Neptune: Voyager Observations of Cloud Motions.
J. Geophys. Res. 96, 18941-18960.

\item[] Lindal, G.\ F., J.\ R. Lyons, D.\ N. Sweetnam, V.\ R. Eshelman, D.\ P. Hinson,
and G.\ L. Tyler 1987. The atmosphere of Uranus: Results of radio occultation measurements
with Voyager 2. J. Geophys. Res. 92, 14987-15001.

\item[] Pearl, J.\ C., and B.\ J. Conrath, R.\ A. Hanel, J.\ A. Pirraglia, and A. Coustenis 1990.
The albedo, effective temperature, and energy balance
of Uranus, as determined from Voyager IRIS data. Icarus 84, 12-28.

\item[] Pearl, J.\ C., and B.\ J. Conrath 1991. The Albedo, Effective Temperature, and Energy Balance
of Neptune, as Determined from Voyager Data. J. Geophys. Res. 18921-18930.

\item[] Pedlosky, J., 1979. Geophysical Fluid Dynamics. Springer-Verlag, New York, 624 pages.

\item[] Rages, K.A., Pollack, J.B., Tomasko, M.G. Doose, L.R.,
  1991. Propertiesw of scatterers in the troposphere and lower
  stratospher of Uranus based on Voyager imaging data. Icarus 89,
  359-376.

\item[] Rages,  K.\ A., H.\ B. Hammel, and A.\ J. Friedson 2004. Evidence for temporal change at
Uranus' south pole. Icarus 172, 548-554.

\item[] Seidelmann, P.\ K., V.\ K. Abalakin, M. Bursa, M.\ E. Davies, C. de Bergh, J.\ H. Lieske, 
J. Oberst, J.\ L. Simon, E.\ M. Standish 2002. Report of the IAU/IAG Working Group on Cartographic
Coordinates and Rotational Elements of the Planets and Satellites: 2000. Celestial Mechanics
and Dynamical Astronomy 82, 83-110.

\item[] Smith, B.A., L.A. Soderblom, R.F. Beebe, D. Bliss,
J.M. Boyce, A. Brahic, G.A. Briggs, R.H. Brown, S.A. Collins,
A.F. Cook II, S.K. Croft, J.N. Cuzzi, G.E. Danielson, M.E. Davies,
T.E. Dowling, D. Godfrey, C.J.  Hansen, C. Harris, G.E. Hunt,
A. P. Ingersoll, T.V. Johnson, R.J. Krauss, H. Masursky, D. Morrison,
T. Owen, J.B. Plescia, J .B.  Pollack, C.C. Porco, K. Rages, C. Sagan,
E.M. Shoemaker, L.A. Sromovsky, C. Stoker, R.G. Strom, V.E.  Suomi,
S.P. Synnot, R.J. Terrile, P. Thomas, W.R. Thompson, and J. Veverka
(1989). Voyager 2 in the Uranian System: Imaging Science Results. {\it
Science \bf 233}, 43-64.

\item[] Sromovsky, L.\ A., S.\ S. Limaye, and P.\ M. Fry 1993. Dynamics of
Neptune's Major Cloud Features. Icarus 105, 110-141.

\item[] Sromovsky, L.\ A., J.\ R. Spencer, K.\ H. Baines, and P.\ M. Fry 2000. Ground-Based
Observations of Cloud Features on Uranus. Icarus 146, 307-311.

\item[]  Sromovsky, L.A., P.M. Fry, K.H. Baines, S.S. Limaye, G.S. Orton,
 and T.E. Dowling 2001.
Coordinated 1996 HST and IRTF Observations of Neptune
and Triton I: Observations, Navigation, and Differential Deconvolution,
{\it Icarus \bf 149}, 416-434.



\item[] Sromovsky, L.\ A., V.\ E. Suomi, J.\ B. Pollack, R.\ J. Krauss, S.\ S. Limaye,
T. Owen, H.\ E. Revercomb, and C. Sagan 1981. Implications of Titan's north-south
brightness asymmetry. Nature 292, 698-702. 

\item[] Sromovsky, L.A., Fry, P.M., Baines, K.H., 2002. The unusual
  dynamics of northern dark spots on Neptune. Icarus 156, 16-36.

\item[] Sromovsky, L.\ A., and P.\ M. Fry 2004. Keck 2 AO Observations of Uranus in 2004.
Bull. Am. Astron. Soc. 36, 1073.

\item[] Sromovsky, L.\ A., and P.\ M. Fry 2005. Cloud Structure on Uranus. Icarus,
 in preparation.

\item[] Stratman, P. W., A.\ P. Showman, T.\ E. Dowling, and L.\ A. Sromovsky 2001. EPIC
Simulations of Bright Companions to Neptune's Great Dark Spots. Icarus
151, 275-285.

\item[] Thompson, D., E. Egami, and M. Sawicki 2001. NIRC-2 The Keck Near-Infrared AO Camera
Pre-ship testing. California Institute of Technology report available at
 http://www2.keck.hawaii.edu/ inst/nirc2/NIRC2Index.html.

\item[] Wallace, L., 1983. The Seasonal Variation of the Thermal Structure of the Atmosphere
of Uranus. Icarus 54, 110-132.

\end{description}

\newpage

\vspace{1.5in}

\newpage

\vspace{0.25in}

\newpage

\newpage


\vspace{0.5in}

\newpage

\newpage

\vspace{1in}



\end{document}